%% file: main.tex
\documentclass[11pt, a4paper, logo, onecolumn, copyright]{googledeepmind}

\usepackage[authoryear, sort&compress, round]{natbib}
\usepackage[ruled,linesnumbered]{algorithm2e}
\usepackage{algpseudocode}
\usepackage{amsfonts}
\usepackage{amsmath}
\usepackage{balance}
\usepackage[table]{xcolor}
\usepackage{booktabs}
\usepackage{comment}
\usepackage[separate-uncertainty,table-align-uncertainty,retain-zero-uncertainty]{siunitx}
\usepackage{tablefootnote}
\usepackage{graphicx}
\usepackage{listings}  
\usepackage{enumitem} 
\definecolor{codegray}{gray}{0.95}
\definecolor{commentgreen}{rgb}{0.13, 0.54, 0.13}
\definecolor{keywordblue}{rgb}{0.13, 0.13, 1}

{}
{}
\definecolor{codegreen}{rgb}{0,0.6,0}
\definecolor{codegray}{rgb}{0.5,0.5,0.5}
\definecolor{codepurple}{rgb}{0.58,0,0.82}
\definecolor{codeblue}{rgb}{0.0,0.0,0.6}
\setlist[enumerate]{nosep, wide, labelwidth=0pt, labelindent=0pt}

\setlist[itemize]{nosep, wide, labelwidth=0pt, leftmargin=0pt, labelindent=0pt}

\title{Discovering Multiagent Learning Algorithms with Large Language Models}

\correspondingauthor{lizun@google.com}


\author[1]{Zun Li}
\author[1]{John Schultz}
\author[1]{Daniel Hennes}
\author[1]{Marc Lanctot}

\affil[1]{Google DeepMind}

\begin{abstract}

\input{abstract}
\end{abstract}

\begin{document}

\maketitle

\input{intro}

\input{preliminaries}
\input{method}
\input{experiments}

\input{related_work}
\input{conclusion}

\input{acknowledgement}

\bibliography{references}

\input{appendix}

\end{document}

%% file: abstract.tex
Much of the advancement in Multi-Agent Reinforcement Learning (MARL) for imperfect-information games has historically depended on the manual, iterative refinement of algorithmic baselines. Recently, evolutionary coding agents powered by Large Language Models (LLMs) have emerged as powerful tools to automate this discovery process. 
In this work, we deploy one of such agentic frameworks, AlphaEvolve, to navigate the design spaces of two distinct game-theoretic paradigms: counterfactual regret minimization (CFR) and policy-space response oracles (PSRO). 
This automated search yielded two algorithms: Volatility-Adaptive Discounted (VAD-) CFR and Smoothed Hybrid Optimistic Regret (SHOR-) PSRO, which are consistently competitive with state-of-the-art human-designed baselines across an 18-game evaluation suite spanning Poker, Goofspiel, Liar's Dice, Blotto, and Battleship variants.
However, because the LLM optimizes for fitness on a specific training set, it often constructs highly synergistic, complex mechanisms tailored to those environments. Through systematic ablation studies, we demonstrate that while these mechanisms are tightly coupled, the true driver of generalization lies in a minimal algorithmic core. By distilling the LLM's discoveries down to their most fundamental principles, we produce two minimal solvers: Warm-started Optimistic Predictive (WOP-)CFR and Projection Matching (PM-)PSRO. These distilled versions achieve superior performance on generalization with greatly reduced structural complexity, providing a clear methodology for using LLMs in algorithmic discovery.

%% file: intro.tex
\section{Introduction}
The field of Multi-Agent Reinforcement Learning has achieved remarkable milestones in recent years, reaching superhuman performance in domains ranging from Poker to real-time strategy games. These advances have been driven by a diverse array of methods, including game-theoretic regret minimization~\citep{Brown2019Pluribus} and population-based league training~\citep{Vinyals2019AlphaStar}.
The practical performance of these algorithms relies heavily on specific structural choices—such as how regret is discounted over time or how a specific equilibrium solution concept is derived.
Historically, the refinement of these choices has been a largely manual endeavor. 
Researchers must rely on intuition and trial-and-error to navigate a vast combinatorial space of potential update rules, often defaulting to mathematically tractable heuristics (e.g., linear averaging or fixed discounting) that may not be optimal.

In this work, we propose to overcome this limitation by automating the design process itself with Large Language Models (LLMs). 
We apply AlphaEvolve~\citep{novikov2025alphaevolve}, a distributed evolutionary system powered by LLMs, to the domain of multi-agent learning. Unlike traditional hyperparameter optimization or genetic programming, AlphaEvolve leverages the code-generation capabilities of LLMs to perform semantic evolution. By treating the algorithm’s source code as the genome, the system uses LLMs to act as intelligent genetic operators—performing mutation to rewrite logic, introduce new control flows, and inject novel symbolic operations. This allows to search beyond simple parameter tuning and discover new non-intuitive mechanisms for equilibrium finding.
This automated search initially yielded two highly performant variants: Volatility-Adaptive Discounted (VAD-)CFR and Smoothed Hybrid Optimistic Regret (SHOR-)PSRO, both consistently competitive with state-of-the-art human-designed baselines across an 18-game suite covering five game families.

However, our analysis reveals that evolutionary agents can layer complex, specialized logic over foundational discoveries to maximize fitness on specific training games. 
For example, the namesake volatility-tracking in VAD-CFR operates in tight conjunction with regret boosting; while highly effective on training games, this complexity is unnecessary for generalization. To isolate the fundamental principles driving performance, we conduct systematic train-test split ablation studies.
By distilling the algorithm down to its most important core components, we introduce two minimal, mathematically pure solvers: Warm-started Optimistic Predictive (WOP-)CFR and Projection Matching (PM-)PSRO. These distilled variants achieve superior generalization while reducing the heuristic complexity present in the raw discoveries.

Our main contributions are threefold: (1) We demonstrate that LLM-driven evolutionary search can navigate the design spaces of two distinct multi-agent learning paradigms (CFR and PSRO) to discover algorithms — VAD-CFR and SHOR-PSRO, which are consistently competitive with state-of-the-art human-designed baselines across an 18-game evaluation suite. (2) Through rigorous train-test ablations, we identify how LLMs construct specialized mathematical complexity to overfit algorithmic structures to their training distributions. (3) We introduce WOP-CFR and PM-PSRO—succinct, highly interpretable algorithms that strip away over-specialized complexity to achieve competitive empirical performance compared to their raw LLM counterparts.

%% file: preliminaries.tex
\section{Game Theoretic Preliminaries}

We formulate our problem within the framework of \textbf{Extensive-Form Games (EFGs)} with imperfect information, which models sequential interactions involving multiple agents and hidden information.

\subsection{Extensive-Form Games and Exploitability}

An $N$-player extensive-form game is $\Gamma = \langle \mathcal{N}, \mathcal{H}, \mathcal{Z}, \mathcal{A}, u, \mathcal{I} \rangle$, where $\mathcal{N}=\{1,\dotsc,N\}$ denotes the set of players~\citep{ShohamLeytonBrown2008}. $\mathcal{H}$ is the set of all possible histories (sequences of actions), where $\mathcal{Z} \subseteq \mathcal{H}$ represents terminal histories.
For $i \in \mathcal{N}$, let $\mathcal{H}_i \subseteq H$ represent the subset of histories where player $i$ acts.
For any non-terminal history $h$, $\mathcal{A}(h)$ is the set of legal actions. The utility function $u_i: \mathcal{Z} \rightarrow \mathbb{R}$ assigns a payoff to player $i$ at a terminal node.
Crucially, imperfect information is modeled via \textbf{Information Sets} $I \in \mathcal{I}_i$. Specifically, $\mathcal{I}_i$ partitions the histories $\mathcal{H}_i$ such that player $i$ cannot distinguish between histories $h, h' \in I$ (e.g., due to hidden cards).
It is required that $\mathcal{A}(h) = \mathcal{A}(h')$ for all $h, h' \in I$, so for simplicity we denote the legal actions at $I$, $\mathcal{A}(I)$.
A \textbf{strategy} (or policy) $\sigma_i(I)$ assigns a probability distribution over actions $a \in \mathcal{A}(I)$ for each information set. A strategy profile $\sigma = (\sigma_1, \dotsc, \sigma_N)$ determines the expected utility $u_i(\sigma)$.
A \textbf{Nash Equilibrium (NE)} is a strategy profile $\sigma^*$ such that no player can increase their utility by deviating unilaterally: $u_i(\sigma_i^*, \sigma_{-i}^*) \ge u_i(\sigma_i', \sigma_{-i}^*) \quad \forall \sigma_i', \forall i \in \mathcal{N}$.
To measure the performance of our evolved algorithms, we use \textbf{Exploitability}. The exploitability of a strategy profile $\sigma$ is the average of the incentives for players to deviate to their Best Response (BR): $\text{Expl}(\sigma) = \frac{1}{N}\sum_{i \in \mathcal{N}} \left( \max_{\sigma_i'} u_i(\sigma_i', \sigma_{-i}) - u_i(\sigma) \right)$.
In small or medium-sized games, we can compute this value exactly by traversing the full game tree.

\subsection{Counterfactual Regret Minimization (CFR)}

CFR is an iterative algorithm that minimizes \textbf{counterfactual regret}~\citep{Zinkevich2007CFR}. It decomposes the global regret minimization problem into independent local regret minimization problems at each information set. Let $\pi^\sigma(h)$ be the probability of reaching history $h$ under strategy profile $\sigma$. We define $\pi^\sigma_{-i}(h)$ as the contribution of all players \textit{except} $i$ (including chance) to reaching $h$. The \textbf{counterfactual value} of player $i$ reaching information set $I$ and playing action $a$ is the expected utility given that player $i$ played to reach $I$: $
    v_i(\sigma, I, a) = \sum_{h \in I} \pi_{-i}^\sigma(h) \sum_{z \in \mathcal{Z}, h \sqsubset z} \pi^\sigma(z~|~h, a) u_i(z)$.
The \textbf{instantaneous counterfactual regret} for not playing action $a$ at iteration $t$ is the difference between the counterfactual value of action $a$ and the expected value at $I$: $r^t_i(I, a) = v_i(\sigma^t, I, a) - \sum_{a' \in \mathcal{A}(I)} \sigma^t_i(I, a') v_i(\sigma^t, I, a')$.
Standard CFR accumulates these values linearly over iterations $T$:
\begin{equation}
    R^T_i(I, a) = \sum_{t=1}^T r^t_i(I, a)\label{eq:cfr_regret_acc}
\end{equation}
The current policy $\sigma^{t+1}_i$ is derived from accumulated regret, typically using \textbf{Regret Matching (RM)}, which assigns probabilities proportional to positive regret:
\begin{equation}
    \sigma^{t+1}_i(I, a) = \frac{\max(R^t_i(I, a), 0)}{\sum_{a'} \max(R^t_i(I, a'), 0)}\label{eq:rm}
\end{equation}

The strategy $\sigma^t_i$ at any single iteration may not converge to NE. Instead, CFR outputs the \textbf{average strategy} $\bar{\sigma}^T_i$, computed by weighting the iteration strategy $\sigma^t_i$ by the player's contribution to the reach probability $\pi_i^{\sigma^t}(I)$.
Several variants modify these update rules. For example, \textbf{CFR+}~\citep{tammelin2014solving} replaces the regret accumulation $R^t_i$ with floor bounding $\max(R^{t-1}_i + r^t_i, 0)$ and uses linear averaging weights ($w_t = t$) for the average policy. Our work uses AlphaEvolve to search for optimal variations of these accumulation and derivation functions.

\subsection{Policy Space Response Oracles (PSRO)}

PSRO~\citep{Lanctot2017PSRO,bighashdel2024policy} acts as a meta-solver that generalizes the Double Oracle algorithm~\citep{mcmahan2003planning}. It operates on a higher level of abstraction called the \textbf{Meta-Game} (or Empirical Game~\citep{wellman2025empirical}).
PSRO maintains a population of policies $\Pi_i = \{\sigma_i^1, \dots, \sigma_i^k\}$ for each player. The meta-game is represented by a payoff tensor $M$, where entries $M_i^{j_1\dotsc j_N} = u_i(\sigma^{j_1}_1,\dotsc, \sigma^{j_N}_N)$ correspond to the expected utility of player $i$ when policies from the population are pitted against each other.
At each epoch $k$, the algorithm performs three steps:
\begin{enumerate}
    \item \textbf{Meta-Strategy Solver (MSS) at training-time:} A solver computes a meta-strategy $\phi_i$ (a probability distribution over the population $\Pi_i$). Common solvers include \textbf{Uniform} $\phi_i^{j_i} = 1/|\Pi_i|$ and \textbf{Nash} $\phi$ which is a Nash Equilibrium of the current meta-game $M$.
    \item \textbf{Oracle (Best Response):} A new policy $\sigma_i^{k+1}$ is trained via Reinforcement Learning (or exact solving) to be a Best Response to the opponent's meta-strategy: $\sigma_i^{k+1} \in {\arg\max}_{\sigma_i}\mathbb{E}_{\sigma_{-i}\sim \phi_{-i}}\left[u_i(\sigma_i, \sigma_{-i})\right]$.
    In this work, we utilize an exact oracle that computes the optimal best response to the meta-strategy, isolating the performance of the meta-solver from the variance of reinforcement learning.
    \item \textbf{Expansion:} The new policy is added to the population $\Pi_i \leftarrow \Pi_i \cup \{\sigma_i^{k+1}\}$, and the payoff tensor $M$ is updated. In this work, we calculate the exact expected payoff value for each entry $M_i^{j_1\dotsc j_N}$, thereby eliminating the stochastic noise associated with Monte Carlo sampling.
\end{enumerate}

To quantify performance, exploitability must be calculated using a meta-strategy distribution over the current population. Standard PSRO typically relies on a single, fixed meta-solver (e.g., Uniform or Nash) for both the training-time oracle generation and the evaluation-time metric calculation. 
By exposing both the training and evaluation solvers as completely separate, evolvable classes, we allow AlphaEvolve to explore whether unified mathematical frameworks or decoupled, asymmetric schedules yield optimal learning dynamics.

%% file: method.tex
\section{Automating Algorithm Discovery for Multiagent Learning via LLMs}\label{sec:methods}
We utilize \textbf{AlphaEvolve}~\citep{novikov2025alphaevolve}, an evolutionary coding agent that leverages Large Language Models (LLMs) to automate the design of multi-agent learning algorithms.
We apply this framework to two distinct paradigms: Regret Minimization (CFR) and Population-based Training (PSRO).

\subsection{The AlphaEvolve Framework}

AlphaEvolve combines LLM code generation with evolutionary selection. We initialize a population $\mathbf{P}$ with the standard implementation of the baseline algorithm (e.g., CFR or Uniform PSRO). In each generation, a parent $A \in \mathbf{P}$ is sampled by fitness; AlphaEvolve supports multi-objective scoring, so one objective is randomly selected per round and parents are sampled in favor of high values on that objective. The parent's source code is fed to an LLM (Gemini 2.5 Pro~\citep{comanici2025gemini}) with the prompt \textit{"Modify the following code to improve fitness (reduce exploitability),"} yielding a candidate $A'$. The candidate is executed on a set of proxy games and scored by final negative exploitability; valid candidates are added to $\mathbf{P}$ and the loop repeats.

\subsection{Evolving Regret Minimization Code}

To discover novel variants of Counterfactual Regret Minimization (CFR) without constraining the evolutionary agent to predefined mathematical structures, we expose the algorithm's core update loop as the search space. Rather than prompting the LLM to write an entire solver from scratch, we extract the CFR framework into three distinct, evolvable Python classes with fixed function signatures in Listing~\ref{lst:cfr_skeleton} in the Appendix:
\begin{enumerate}
    \item \texttt{RegretAccumulator}: Dictates how instantaneous counterfactual regrets are processed, discounted, or bounded before being added to the historical cumulative regret.
    \item \texttt{PolicyFromRegretAccumulator}: Defines the projection mechanism (e.g., Regret Matching) used to derive the current iteration's strategy from the accumulated regrets.
    \item \texttt{PolicyAccumulator}: Determines how the current iteration's policy is weighted and absorbed into the final average strategy used for evaluation.
\end{enumerate}

This search space is expressive enough to encompass the entire family of known CFR variants as special cases; for instance, standard CFR uses eq~(\ref{eq:cfr_regret_acc}) for $\texttt{RegretAccumulator}$ and eq~(\ref{eq:rm}) for $\texttt{PolicyFromRegretAccumulator}$.

\subsection{Evolving Meta-Strategy Solvers Code}

For PSRO, we evolve python classes \texttt{TrainMetaStrategySolver} for generating the mixed strategies used during the oracle training phase, and \texttt{EvalMetaStrategySolver} for computing a strategy profile over population to report exploitability.
This interface (Listing~\ref{lst:solvers} in the Appendix) supports the representation of all standard baselines as special cases: for example, standard double oracle algorithm solves for a Nash equilibrium in the $\texttt{get\_meta\_strategy}$ method for both solver classes.
By exposing the training and evaluation steps as two completely separate, evolvable Python classes, we grant the LLM the maximum degree of freedom to discover dynamic, asymmetric schedules between training and evaluation.

\subsection{Meta-Training Objective}
We manually select a set $\mathcal{G}_{train}$ of training games for AlphaEvolve to compute $|\mathcal{G}_{train}|+1$ fitness scores.
These are the negative exploitability $-\text{Expl}(A(g)_K)$ of the final strategy profile after $K$ iterations for each $g\in\mathcal{G}_{train}$, as well as their average
$$- \frac{1}{|\mathcal{G}_{train}|} \sum_{g \in \mathcal{G}_{train}} \text{Expl}(A(g)_K).$$
Here $A(g)_K$ denotes the strategy produced by algorithm $A$ on game $g$ at iteration $K$.
The reported algorithms are selected based on their average scores.

%% file: experiments.tex
\section{Experimental Evaluation}
\label{sec:experiments}
\begin{figure*}[t!]
    \centering
    \includegraphics[width=1\textwidth]{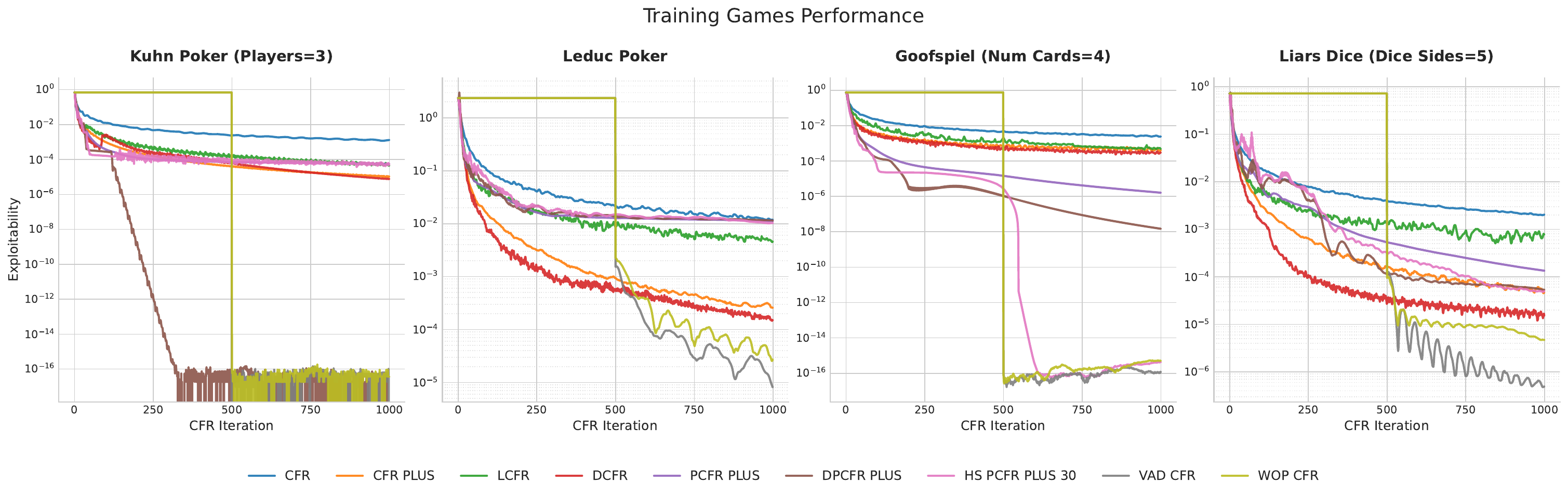}
    \centering
    \includegraphics[width=1\textwidth]{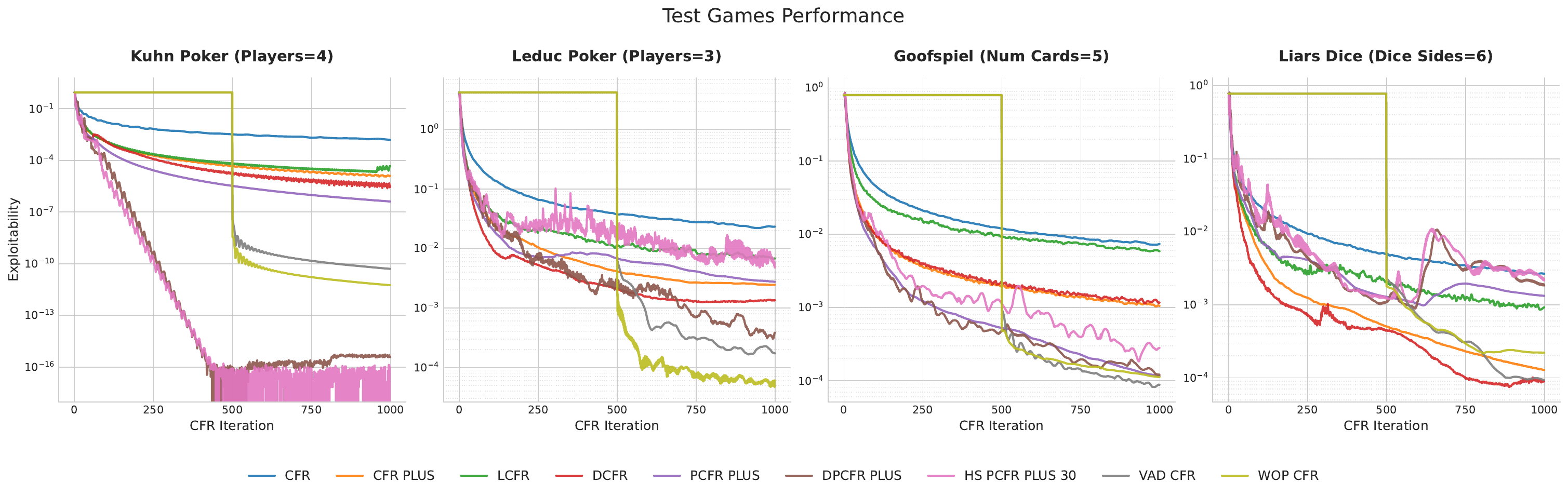}
    \caption{\textbf{CFR variants performances.}}
    \label{fig:results_cfr}
\end{figure*}

\subsection{Experimental Setup}\label{sec:setup}
To test the robustness and generalizability of the algorithms we discover, we adopted a rigorous evaluation protocol involving two distinct sets of games. 
The algorithm's architecture and hyperparameters were developed and tuned on a \textit{Training Set} $\mathcal{G}_{train}$ of four games. 
For both CFR and PSRO discoveries, we choose this set to be 3-player Kuhn Poker, 2-player Leduc Poker, 4-card Goofspiel, and 5-sided Liars Dice.
Subsequently, the fixed algorithm was evaluated on a \textit{Test Set} $\mathcal{G}_{test}$.
In the main body of this paper, we present results on four larger and more difficult games as a representative subset of test games: 4-player Kuhn Poker, 3-player Leduc Poker, 5-card Goofspiel, and 6-sided Liar's Dice.
The results of a full-sweep of four training games and fourteen test games are provided in Appendix~\ref{sec:full_results}.
We utilized the OpenSpiel~\citep{lanctot2019openspiel} framework for all experiments.
For all experiments we use a uniform policy as the initial point.
Our implementation of AlphaEvolve is backboned by Gemini 2.5 pro~\citep{comanici2025gemini}.
All evaluated algorithms are fully deterministic and metrics are computed exactly by exhaustively traversing the game-tree.

\subsection{Ablation Methodology}\label{sec:ablation}
AlphaEvolve may exhibit a tendency to over-engineer solutions, interleaving profound algorithmic discoveries with symbolic complexity. To rigorously isolate the true engines of performance from these over-specialized heuristics, we introduce a comprehensive algorithmic auditing pipeline via train-test split ablations.
We adopt the Interquartile Mean (IQM)~\citep{agarwal2021deep,oh2025discovering}, a robust statistical aggregator that is highly resistant to outlier domains.
For a given set of games $\mathcal{G}$, let $\text{Expl}_{\text{raw}}^{(g)}$ and $\text{Expl}_{\text{abl}}^{(g)}$ denote the final exploitability of the raw LLM algorithm and its ablated counterpart on game $g\in\mathcal{G}$, respectively. We first define the \emph{log-improvement score} $s_g$ for each game: $s_g = \log_{10}\left(\text{Expl}_{\text{raw}}^{(g)}\right) - \log_{10}\left(\text{Expl}_{\text{abl}}^{(g)}\right)$.
Under this formulation, a positive score $s_g > 0$ indicates that the ablated (simplified) algorithm achieves lower exploitability than the raw discovery, identifying the removed component as an over-parameterized generalization penalty. Conversely, a negative score $s_g < 0$ indicates a degradation in performance. 
To aggregate these scores across the diverse game tasks, the IQM computes the mean of the central 50\% of the scores, systematically trimming the top 25\% and bottom 25\% of extreme variations.
To isolate generalization effects, we report the IQM across three distinct splits: Train IQM (on $\mathcal{G}_{train}$) serves as our primary diagnostic for overfitting, Test IQM (on $\mathcal{G}_{test}$) acts as our ultimate metric for generalization, and Total IQM provides a global measure of algorithmic robustness across all evaluated domains.

\begin{table*}[t!]
\centering
\caption{\textbf{Ablation of VAD-CFR Components.} Results are reported as IQM Log-Improvement Scores ($s_g$). A negative score of an individual ablation indicates that removing the component degrades performance.}
\label{tab:vad_cfr_audit}
\begin{tabular}{lccc}
\toprule
\textbf{Variant} & \textbf{Total IQM $\boldsymbol{s_{total}}$} & \textbf{Train IQM $\boldsymbol{s_{train}}$} & \textbf{Test IQM $\boldsymbol{s_{test}}$} \\ \midrule
\textbf{WOP-CFR (Distilled)} & \textbf{-0.073} & \textbf{-0.802} & \textbf{+0.119} \\
\midrule
\textit{Individual VAD-CFR Ablations:} & & &  \\
\texttt{no\_volatility} & -0.120 & -0.842 & -0.029  \\
\texttt{no\_boost} & -0.275 & -0.954 & -0.204  \\
\texttt{no\_predication} & -1.584 & \textbf{-7.029} & -0.905  \\
\texttt{no\_asym\_discount} & -1.637 & -4.192 & -1.065  \\
\texttt{no\_warmstart} & \textbf{-2.733} & -5.256 & \textbf{-2.003}  \\
\bottomrule
\end{tabular}
\end{table*}

\subsection{Experimental Evaluation: Discovering and Distilling State-of-the-Art CFR}\label{sec:cfr_results}
The evolutionary search over the CFR design space yielded a remarkably sophisticated algorithm we term Volatility-Adaptive Discounted (VAD-)CFR, whose source code is provided in Listing~\ref{lst:vadcfr}.
We benchmarked evolved VAD-CFR against a suite of state-of-the-art regret minimization algorithms: standard CFR~\citep{Zinkevich2007CFR}, CFR+~\citep{tammelin2014solving}, Linear CFR (LCFR), Discounted CFR (DCFR)~\citep{brown2019solving},  Predictive CFR+ (PCFR+)~\citep{farina2021faster}, Discounted Predictive CFR+ (DPCFR+)~\citep{xu2024minimizing}, and a Hyperparameter Schedule-powered PCFR variant HS-PCFR+(30)~\citep{zhang2024faster}. Performance was quantified using exploitability (measured on a logarithmic scale) over a fixed horizon of $K=1000$ iterations.
We use CFR+ as the seed program.
The prompt we use is shown in Listing~\ref{lst:cfr_prompt} in the Appendix.

As shown in Figure~\ref{fig:results_cfr}, VAD-CFR ranks among the top performers across both the training set and the test set.
For 3-player Kuhn Poker, VAD-CFR achieves significantly lower exploitability than all baselines.
For Leduc Poker and 4-card Goofspiel, the algorithm maintains a steeper convergence slope compared to DPCFR+ and other state-of-the-art variants.
In 5-sided Liars Dice, VAD-CFR exhibits robust performance, effectively managing the larger state space through its adaptive discounting and boosting mechanisms.
Results illustrated in test games highlight its generalization.
In 3-player Leduc Poker, VAD-CFR reaches exploitability levels below $10^{-3}$ while most baselines plateau at higher levels.
In 6-sided Liar's Dice, VAD-CFR continues to match established baselines like DCFR, suggesting that its evolved mechanisms for managing regret scaling and noise are highly effective across different tasks.
Overall, VAD-CFR demonstrates efficient convergence rate and generalization across a broad variety of domains: as shown in Figure~\ref{fig:cfr_all_games} in the Appendix, \textbf{compared with baselines, VAD-CFR ranks in the top three on every one of the 18 games, and achieves the lowest exploitability on 15 of them}.

\subsubsection{Identifying Foundational Mechanics}

The raw VAD-CFR discovery (Listing~\ref{lst:vadcfr}) operates as a highly synergistic ecosystem of update rules. To isolate the true engines of performance from the LLM's over-specialized heuristics, we conducted individual feature ablations using our IQM log-improvement score ($s_g$). 
The details of all ablations are in Appendix~\ref{sec:vad_ablations}.

Crucially, removing secondary features such as EWMA volatility tracking (\texttt{no\_volatility}, $s_{train}=-0.842,s_{test}=-0.029$) or instantaneous regret boosting (\texttt{no\_boost}, $s_{train}=-0.954,s_{test}=-0.204$) degrades training fit while having only marginal effects on held-out generalization (Table~\ref{tab:vad_cfr_audit}). 
This indicates that the LLM evolved a tightly coupled system where these secondary features matter for in-distribution synergy but contribute little to generalization.
However, while this synergistic complexity maximizes fitness on the training domains, our audit reveals that the true capacity for generalization rests entirely on three foundational structural shifts. Removing any of these causes a catastrophic collapse in exploitability:

\begin{itemize}
    \item \textbf{Hard Warm-Start ($s_{train}=-5.256, s_{test} = -2.003$):} The most critical feature discovered by the LLM is a strict $500$-iteration suspension of all policy averaging. By enforcing a hard delay, it ensures that early-stage, high-variance exploration noise does not pollute the final cumulative strategy profile. 

    \item \textbf{Extreme Asymmetric Discounting ($s_{train}=-4.192, s_{test} = -1.065$):} While standard DCFR establishes the value of decoupling positive and negative regret discounting, the LLM pushes this to a mathematical extreme. It assigns a negative exponent ($\beta = -0.1$) for negative regrets. Because the discount multiplier is driven by a negative exponent, it rapidly decays toward zero, acting as an almost instantaneous ``forget-gate'' that prevents regret lock-in significantly faster than human-tuned DCFR baselines.

    \item \textbf{Optimistic Prediction ($s_{train}=-7.029, s_{test} = -0.905$):} VAD-CFR abandons standard linear Regret Matching, instead deriving the policy using a non-linear prediction with an exponent of $1.5$ and a decaying optimism factor. This allows the policy to ``anticipate'' opponent shifts while aggressively correcting major strategic deficits.

\end{itemize}

\subsubsection{Distillation: Warm-started Optimistic Predictive (WOP-) CFR}

While single-feature ablations destabilize the solver’s synergy, simultaneously stripping away all secondary heuristics reveals a highly distilled algorithmic core. 
By preserving only the three foundational components identified in the audit, we produce Warm-started Optimistic Predictive (WOP-) CFR (Listing~\ref{lst:wop_cfr}).
Unlike the raw discovery, WOP-CFR can be defined purely through three formal mathematical modifications to the standard CFR framework. 
First, it applies the discovered extreme asymmetric discounting to the accumulated regret $R^t_i(I,a)$:$$R^t_i(I,a) =  d^{(\pm)}_t \cdot R^{t-1}_i(I,a) + r^t_i(I,a)$$ where $d^{(+)}_t = \frac{t^{1.5}}{t^{1.5}+1}$ if $R^{t-1}_i(I,a) \ge 0$ and $d^{(-)}_t = \frac{t^{-0.1}}{t^{-0.1}+1}$ otherwise. Second, it derives the current policy using a non-linear, optimistic predication for the next target:$$\sigma^{t+1}_i(I, a) \propto \max\left(0,\ d^{(\pm)}_{t} \cdot R^{t}_i(I, a) + \frac{1}{1+t/100}\cdot r^t_i(I, a)\right)^{1.5}$$ where $d^{(\pm)}_t$ now depends on the sign of $R^t_i(I,a)$ rather than $R^{t-1}_i(I,a)$.
Finally, it enforces the strict hard warm-start, suspending average policy accumulation entirely for the first $500$ iterations, after which it accumulates using a static polynomial weight $w_t = t^2$. By discarding the LLM's overfitted training heuristics and structurally enforcing these three core mechanics, this distilled variant achieves a \textbf{+0.119 Test IQM} improvement over the raw VAD-CFR discovery.
As shown in Figure~\ref{fig:cfr_all_games} in the Appendix, \textbf{compared with baselines, WOP-CFR ranks in the top three on every one of the 18 games, and achieves the lowest exploitability on 14 of them}.
In certain games like 3-player Leduc or 4-card limited information Goofspiel, it even outperforms VAD-CFR by non-negligible margins.

\begin{figure*}[!t]
    \centering
    \includegraphics[width=1\textwidth]{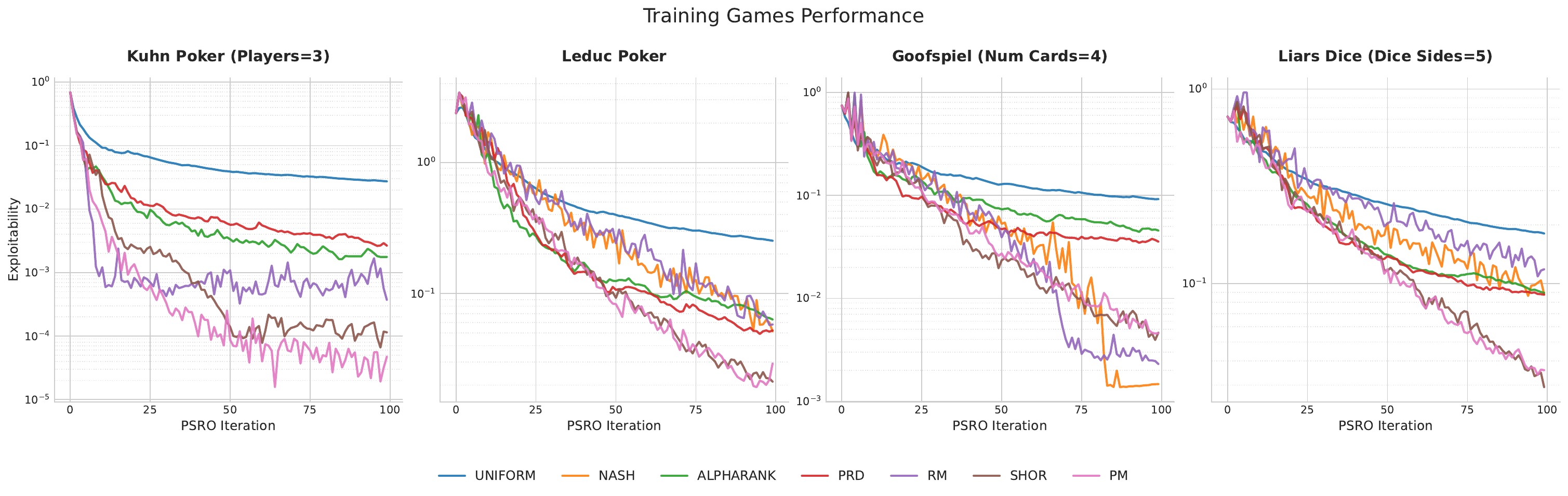}
    \centering
    \includegraphics[width=1\textwidth]{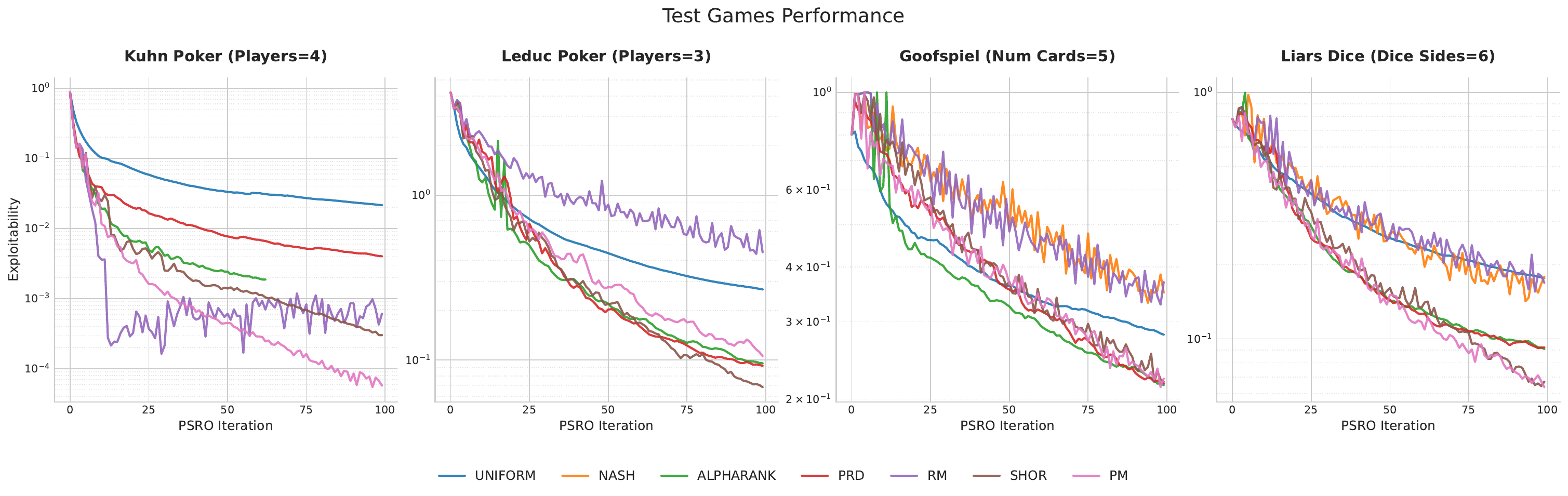}
    \caption{\textbf{PSRO variants performances.}}
    \label{fig:results_psro}
\end{figure*}

\subsection{Experimental Evaluation: Discovering and Distilling State-of-the-Art PSRO}
\label{sec:exp_shorpsro}

Next, we evaluated the performance of the evolved Smoothed Hybrid Optimistic Regret (SHOR)-PSRO algorithm, comparing its ability to reduce exploitability against standard meta-solver baselines. 
We use the exploitability at the $K=$100-th PSRO iteration as the metric.
We employ an \emph{exact best response oracle} via value iteration, where at each state it distributes probability mass uniformly among actions with the same optimal values.
We benchmarked SHOR-PSRO against standard established meta-solver baselines: Uniform, Nash equilibrium via linear program for 2-player games, AlphaRank~\citep{mullergeneralized}, Projected Replicator Dynamics (PRD)~\citep{Lanctot2017PSRO}, and Regret Matching (RM) executed for $10^4$ steps per PSRO iteration.
We use Uniform as the initial program for both solver classes.
The prompt we use is shown in Listing~\ref{lst:psro_prompt} in the Appendix.

As shown in Figure~\ref{fig:results_psro}, SHOR-PSRO ranks among the top performers across both the training set and the test set.
In simpler domains like 3-player Kuhn Poker, SHOR-PSRO achieves exploitability levels ($< 10^{-3}$) significantly faster than PRD or RM.
In 3-player Leduc Poker, the meta-game landscape becomes significantly more chaotic due to multi-agent dynamics. 
Despite this, SHOR-PSRO consistently matches or outperforms the best-performing baselines.
In the most demanding test case, Liar's Dice (6 sides), SHOR-PSRO also demonstrates a clear advantage. 

\subsubsection{Identifying Foundational Mechanics}

The raw SHOR-PSRO algorithm (Listing~\ref{lst:shor-psro}) contains several highly synergistic mechanisms, most notably an optimistic regret matching+ (ORM+) engine, a hybrid blending module that actively interpolates between the ORM+ output and a smoothed best pure strategy, and asymmetric solvers that utilize distinct logic for training versus evaluation.

To isolate the core drivers of performance, we conducted a systematic train-test split ablation study (with exhaustive configurations detailed in Appendix~\ref{sec:psro_ablations}). 
As shown in Table~\ref{tab:exhaustive_psro_ablations}, the audit exposes the fragility of the LLM's asymmetric design and blending schedules. 
First, the result identifies the ORM$^+$ loop as the undeniable foundational engine. Removing the ORM$^+$ loop entirely to rely solely on the smoothed pure strategy (\texttt{no\_blending\_pure}) causes a catastrophic collapse in both training and testing environments ($s_{train}= -2.292,s_{test}=-1.268$). 
In contrast, disabling the hybrid pure-strategy blending (\texttt{no\_blending\_orm}, $s_{train}=-0.069$) and removing the dynamic scheduling (\texttt{no\_annealing}, $s_{train}=-0.104$) have negligible effects on training fit, indicating these components do not even drive in-distribution performance. 
Test IQM corroborates this diagnosis on held-out games: \texttt{no\_blending\_pure} collapses Test IQM by -1.268, while removing the hybrid blending and dynamic scheduling each independently improves Test IQM (+0.020, +0.016), confirming they are overfitted artifacts rather than generalization drivers.

\subsubsection{Distillation: Projection Matching (PM-)PSRO}

To isolate the true engine of performance, we bypass the LLM's entangled heuristics and distill the solver down to its mathematical core: \textbf{Projection Matching (PM-)PSRO} (in Listing~\ref{lst:pgm_plus_psro}). By stripping away annealing schedules, hybrid blending, and complex evaluation logic, PM is defined strictly by three essential mechanics.
First, PM replaces the regret signal in regret matching with a tangent-projected utility. At each inner step $t$, given the expected-utility vector $u_i^t \in \mathbb{R}^{|\Pi_i|}$ over the population of player $i$, we project it onto the tangent space of the probability simplex $\mathcal{T}\Delta$
$$g_i^t = \mathcal{P}_{\mathcal{T}\Delta}(u_i^t) = \left(I - \tfrac{1}{|\Pi_i|}\mathbf{11}^\top\right) u_i^t,$$
i.e., $u_i^t$ centered by its mean. We then run standard regret matching on $g$ in place of $r$ (equations (\ref{eq:cfr_regret_acc})(\ref{eq:rm}) at the normal-form level).
The tangent-projection step is the same geometric operation that defines the continuous-time \textbf{Projection Dynamic}~\cite{lahkar2008projection,sandholm2008projection} in evolutionary game theory literature, but the discrete update we use is RM-style rather than the projection-dynamic rule.
Second, PM scales its inner iteration budget linearly with the empirical-game size, so the solver budget tracks population growth. 
Third, PM uses a unified time-averaged solver for both training and evaluation, since our ablations (Table~\ref{tab:exhaustive_psro_ablations}) found it better than a single last-iterate solver.

Centering utilities by the population mean evaluates each policy strictly relative to the population average; paired with RM's positive bounding, the solver instantly "forgets" below-average strategies, providing geometric immunity to extreme payoff outliers. Combining this core mechanism with the dynamic optimization budget yields a \textbf{+0.059 Test IQM} log-improvement over the raw SHOR-PSRO discovery.
As shown in Figure~\ref{fig:psro_all_games} in the Appendix, \textbf{compared with baselines, both SHOR-PSRO and PM-PSRO rank in the top three on every one of the 18 games, and achieve the lowest exploitability on 12 of them}.
We further validate these findings on 1000 randomly sampled $N$-player $A$-action constant-sum normal-form games per $(N, A)$ configuration (see Appendix~\ref{sec:nfg_results}), where SHOR-PSRO and PM-PSRO dominate all baselines across every $(N, A)$ tested.

%% file: related_work.tex
\section{Related Work}
\label{sec:related_work}
Significant research effort has been dedicated to improving the convergence speed of CFR through specific weighting schemes and regret targets. 
Notable variants include CFR+~\citep{tammelin2014solving}, 
strategy-based warm starting~\citep{brown2016strategy},
Discounted CFR (DCFR)~\citep{brown2019solving}, and Predictive CFR+ (PCFR+)~\citep{farina2021faster}; these were derived through human intuition over a vast design space.
Policy Space Response Oracles (PSRO)~\citep{bighashdel2024policy} generalizes the Double Oracle algorithm~\citep{mcmahan2003planning} by iteratively expanding a population of policies via exact or reinforcement learning best responses. 
While PSRO rests on solid theoretical ground~\citep{zhang2024exponential}, its practical convergence and population-quality remain open challenges ~\citep{bighashdel2024policy}.

Research on automating machine learning algorithm design has been evolving for both neural-based approaches and symbolic-based approaches. 
Meta-learning approaches such as meta-RL \citep{xu2018meta,oh2025discovering} and meta-learning optimizers \citep{metz2019understanding} have parameterized update rules using neural networks to optimize learning dynamics.
For symbolic ML discovery, a foundational work in this domain is AutoML-Zero \citep{real2020automl}, which demonstrated that complete machine learning algorithms could be evolved from scratch using basic mathematical operations.
Subsequent work applies program search to discover optimizers~\citep{chen2023symbolic} and symbolic RL algorithms~\citep{coevolving}, and LLM-driven search to discover isolated preference-optimization losses~\citep{lu2024discovering}.
Our contribution differs along two axes: we evolve structured algorithmic skeletons with multiple interacting components rather than a single loss or update rule, and we introduce a formal train-test ablation methodology to distill raw discoveries into minimal solvers — whereas prior work either ships the raw artifact~\citep{lu2024discovering,coevolving} or applies one-off manual simplification~\citep{chen2023symbolic}.

Within the specific domain of multi-agent learning, there has been prior work on automating algorithm design \citep{fengneural, xu2022autocfr, xu2024dynamic,Sychrovsky24}. 
However, these approaches often faced limitations: they either operated within a relatively constrained search space or relied on neural parameterizations that hindered interpretability. 
Our work builds directly on AlphaEvolve \citep{novikov2025alphaevolve}, utilizing Large Language Models (LLMs) to perform semantic mutation on interpretable code, bridging the gap between expressive neural meta-learning and symbolic discovery.
This approach has already shown great success in the field of math~\citep{georgiev2025mathematical} and combinatorial algorithms~\citep{nagda2025reinforced}.

%% file: conclusion.tex
\section{Conclusion}\label{sec:conclusion}
We applied LLM-driven evolutionary search to automate the design of multi-agent learning algorithms in two paradigms — CFR and PSRO. The raw discoveries VAD-CFR and SHOR-PSRO are consistently competitive with state-of-the-art human-designed baselines across an 18-game evaluation suite. Through systematic train-test ablations, we distilled both into minimal solvers, WOP-CFR and PM-PSRO, that retain or improve generalization with substantially reduced complexity.
Our results suggest LLM-driven evolutionary search is most useful as a proposal mechanism embedded in a human-in-the-loop pipeline: the LLM proposes candidate combinations of mechanisms, train-test ablation surfaces which generalize, and the researcher distills the survivors. Future work will extend this pipeline to deep reinforcement learning agents and cooperative general-sum games.

\paragraph{Limitations.} Our results have several caveats. (i) The reported algorithms reflect a single AlphaEvolve trajectory; we do not characterize variance over independent evolutionary runs, and the stability of the distilled components across runs remains empirically unverified, although each is well-motivated independent of any single run. 
(ii) Our headline IQM is dominated by within-family generalization across train/test sets that share three game families; out-of-family generalization (Blotto, Battleship) is scoped to what Appendix~\ref{sec:full_results} shows.
(iii) Distillation requires human judgment to identify which components to retain or simplify; the pipeline is human-in-the-loop, not fully autonomous. (iv) We provide no formal convergence guarantees. 
PM-PSRO uses tangent-projected utilities as the regret signal in standard RM, which shares a geometric step with the projection dynamic~\cite{lahkar2008projection} but is not equivalent to it; WOP-CFR's components are inspired by DCFR~\cite{brown2019solving} and PCFR+~\cite{farina2021faster} but the specific combination is empirically rather than theoretically motivated. 
(v) We only compare with previous tabular-based baselines and do not include deep CFR~\cite{brown2019deep} or neural auto-curricula PSRO~\cite{fengneural}; claims of competitive performance are scoped to the baselines tested. 
(vi) Exact exploitability restricts evaluation to games with less than 350K information states; scaling behavior and dependence on the specific LLM backbone and prompt design remain open.

%% file: acknowledgement.tex
\section*{Acknowledgements}
We thank Ian Gemp and Xidong Feng for their valuable discussions and comments on this work.

%% file: appendix.tex
\newpage

\section{Appendix}
\label{sec:appendix}

\subsection{Code skeleton}\label{sec:skeleton}

\begin{lstlisting}[caption={The Python CFR code skeleton used as the search space for AlphaEvolve. The highlighted methods \texttt{update\_accumulate\_regret}, \texttt{get\_updated\_current\_policy}, and \texttt{update\_accumulate\_policy} represent the evolvable components of the CFR algorithm.}, label={lst:cfr_skeleton}]
class RegretAccumulator:
  """A class that updates cumulative regret at an information set."""

  def update_accumulate_regret(self, info_state_node, iteration_number, cfr_regrets):
    """
    Args:
      info_state_node: Data structure with cumulative_regret and cumulative_policy.
      iteration_number: Current CFR iteration.
      cfr_regrets: Counterfactual regrets (not yet added to node).
    Returns:
      Updated cumulative regret dictionary for each action.
    """
    ...

class PolicyFromRegretAccumulator:
  """A class that derives the current policy from regret."""

  def get_updated_current_policy(self, info_state_node, iteration_number, cfr_regrets, previous_policy):
    """
    Args:
      info_state_node: Data structure with cumulative_regret.
      iteration_number: Current CFR iteration.
      cfr_regrets: Counterfactual regrets (already added to node).
      previous_policy: Previous policy at this info set.
    Returns:
      Updated current policy dictionary.
    """
    ...

class PolicyAccumulator:
  """A class that updates the average policy during tree traversal."""

  def update_accumulate_policy(self, info_state_node, iteration_number, info_state_policy, cfr_regrets, reach_prob, counterfactual_reach_prob):
    """
    Args:
      info_state_node: Data structure with cumulative_policy.
      iteration_number: Current CFR iteration.
      info_state_policy: Current policy at this info set.
      cfr_regrets: Counterfactual regrets (already added to node).
      reach_prob: Probability of reaching current history (player's contribution).
      counterfactual_reach_prob: Probability of reaching current history (opponents' contribution).
    Returns:
      Updated cumulative policy dictionary.
    """
    ...
\end{lstlisting}

\begin{lstlisting}[caption={The Python PSRO code skeleton used as the search space for AlphaEvolve. The highlighted methods \texttt{TrainMetaStrategySolver} and \texttt{EvalMetaStrategySolver} represent the evolvable components of the PSRO algorithm.}, label={lst:solvers}]
class TrainMetaStrategySolver:
  """Returns meta strategies to train against in PSRO."""

  def get_meta_strategy(self, game, policy_sets, meta_games):
    """Returns meta strategies to train against in policy-space response oracles.

    Args:
      game: The pyspiel game object.
      policy_sets: A list of lists of policies, one list per player.
        policy_sets[p][i] is player p's i-th policy. len(policy_sets[p]) ==
        meta_games[0].shape[p].
      meta_games: A list of n-dimensional numpy arrays, one per player. Each
        array has shape (num_strats_p0, num_strats_p1, ..., num_strats_pn-1) and
        meta_games[p][i0, i1, ..., in-1] is the payoff of player p when player k
        chooses strategy ik.

    Returns:
      A list of mixed-strategies, one for each player. Each mixed strategy is
      a list of non-negative weights (not necessarily normalized). It is used
      to train best response against.
    """
    ...

class EvalMetaStrategySolver:
  """Returns meta strategies for evaluation in PSRO."""

  def get_meta_strategy(self, game, policy_sets, meta_games):
    """Returns meta strategies for evaluation in policy-space response oracles.

    Args:
      game: The pyspiel game object.
      policy_sets: A list of lists of policies, one list per player.
        policy_sets[p][i] is player p's i-th policy. len(policy_sets[p]) ==
        meta_games[0].shape[p].
      meta_games: A list of n-dimensional numpy arrays, one per player. Each
        array has shape (num_strats_p0, num_strats_p1, ..., num_strats_pn-1) and
        meta_games[p][i0, i1, ..., in-1] is the payoff of player p when player k
        chooses strategy ik.

    Returns:
      A list of mixed-strategies, one for each player. Each mixed strategy is
      a list of non-negative weights (not necessarily normalized). It is used
      for evaluation of the current PSRO policies. E.g., computing
      exploitability.
    """
    ...
\end{lstlisting}

\subsection{Source code of discovered algorithms}\label{sec:source}

\begin{lstlisting}[
caption={VAD-CFR},
language=Python, 
label=lst:vadcfr,
]
class RegretAccumulator:
  """A class that updates cumulative regret using Adaptive Discounting
  with separate discounting for positive and negative regrets, and instantaneous regret boosting.
  """

  @staticmethod
  def _calculate_adaptive_params(
      iteration_number,
      cfr_regrets,
      base_alpha,
      base_beta,
      volatility_sensitivity,
      max_expected_instantaneous_regret,
      ewma_decay_factor,
      current_ewma_magnitude,
  ):
    """Calculates adaptive discounting parameters for a given iteration.
    This static method centralizes the logic for EWMA volatility, adaptive
    alpha/beta, and discount factors to ensure consistency across components.
    """
    t_plus_one = float(iteration_number + 1)
    
    instantaneous_regret_magnitude = max(
        (abs(r) for r in cfr_regrets.values()), default=0.0
    )

    if iteration_number == 0:
        projected_ewma = instantaneous_regret_magnitude
    else:
        projected_ewma = (
            ewma_decay_factor * instantaneous_regret_magnitude +
            (1.0 - ewma_decay_factor) * current_ewma_magnitude
        )
    
    if max_expected_instantaneous_regret > 0:
        normalized_volatility = min(1.0, projected_ewma / max_expected_instantaneous_regret)
    else:
        normalized_volatility = 0.0

    effective_alpha = max(0.1, base_alpha - volatility_sensitivity * normalized_volatility)
    effective_beta = base_beta - volatility_sensitivity * normalized_volatility
    effective_beta = min(effective_alpha, effective_beta)

    discount_factor_positive = (t_plus_one**effective_alpha) / (t_plus_one**effective_alpha + 1.0)
    discount_factor_negative = (t_plus_one**effective_beta) / (t_plus_one**effective_beta + 1.0)
    
    return projected_ewma, normalized_volatility, discount_factor_positive, discount_factor_negative

  def __init__(self, base_alpha=1.5, base_beta=-0.1, volatility_sensitivity=0.5,
               max_expected_instantaneous_regret=2.0, instantaneous_regret_boost_factor=1.1,
               ewma_decay_factor=0.1, negative_regret_cap=-20.0):
    """Initializes the regret accumulator with adaptive discounting parameters.

    Args:
      base_alpha: The baseline exponent for discounting positive cumulative regrets.
      base_beta: The baseline exponent for discounting negative cumulative regrets.
      volatility_sensitivity: Controls how strongly the instantaneous regret
        magnitude influences the adaptive alpha/beta. A higher value means the
        exponents will be more reduced by high volatility.
      max_expected_instantaneous_regret: An estimate of the maximum possible
        instantaneous regret magnitude, used for normalizing the volatility.
      instantaneous_regret_boost_factor: Boost factor for positive instantaneous regrets.
        A factor > 1.0 makes the algorithm more reactive to current good actions.
      ewma_decay_factor: Decay factor for the EWMA of instantaneous regret magnitude.
      negative_regret_cap: The minimum value for cumulative regret, to prevent
        regret lock-in and improve adaptability.
    """
    self._base_alpha = base_alpha
    self._base_beta = base_beta
    self._volatility_sensitivity = volatility_sensitivity
    self._max_expected_instantaneous_regret = max_expected_instantaneous_regret
    self._instantaneous_regret_boost_factor = instantaneous_regret_boost_factor
    self._ewma_decay_factor = ewma_decay_factor
    self._negative_regret_cap = negative_regret_cap
    self._ewma_instantaneous_regret_magnitude = 0.0

  def update_accumulate_regret(
      self, info_state_node, iteration_number, cfr_regrets
  ):
    """Updates cumulative regret for each action at an information set.
    Cumulative regrets are now signed.

    Args:
      info_state_node: a data structure corresponding to an information set with
        cumulative_regret and cumulative_policy stored.
      iteration_number: the current CFR iteration (0-indexed).
      cfr_regrets: the instantaneous counterfactual regrets of the current policy at the
        current information set. cfr_regrets haven't been added to info_state_node.cumulative_regret.
    Returns:
      updated cumulative regret for each action at the current information set (signed).
    """
    # Centralize adaptive parameter calculation to ensure consistency and reduce redundancy.
    (
        self._ewma_instantaneous_regret_magnitude,
        _,  # normalized_volatility is not used here
        discount_factor_positive,
        discount_factor_negative,
    ) = RegretAccumulator._calculate_adaptive_params(
        iteration_number=iteration_number,
        cfr_regrets=cfr_regrets,
        base_alpha=self._base_alpha,
        base_beta=self._base_beta,
        volatility_sensitivity=self._volatility_sensitivity,
        max_expected_instantaneous_regret=self._max_expected_instantaneous_regret,
        ewma_decay_factor=self._ewma_decay_factor,
        current_ewma_magnitude=self._ewma_instantaneous_regret_magnitude,
    )

    updated_cumulative_regret = {}
    for action in cfr_regrets:
      old_regret = info_state_node.cumulative_regret[action]
      
      instantaneous_regret_component = cfr_regrets[action]
      if instantaneous_regret_component > 0:
          instantaneous_regret_component *= self._instantaneous_regret_boost_factor

      # Apply different discount factors based on the sign of the old regret.
      if old_regret >= 0:
          discounted_old_regret = discount_factor_positive * old_regret
      else:
          discounted_old_regret = discount_factor_negative * old_regret

      new_regret = discounted_old_regret + instantaneous_regret_component
      
      # Cap the negative regret to prevent lock-in and improve adaptability.
      new_regret = max(self._negative_regret_cap, new_regret)

      # Crucially, regrets are NOT clipped to zero here. They can be negative.
      updated_cumulative_regret[action] = new_regret
      
    return updated_cumulative_regret

class PolicyFromRegretAccumulator:
  """A class that obtains a current policy from a consistent optimistic projection of regrets.
  It aligns the policy generation with the adaptive, asymmetric, and boosted logic from RegretAccumulator.
  """

  def __init__(self,
               initial_optimism_factor=1.0,
               optimism_decay_factor=100.0,
               positive_policy_exponent=1.5,
               base_alpha=1.5,
               base_beta=-0.1,
               volatility_sensitivity=0.5,
               max_expected_instantaneous_regret=2.0,
               instantaneous_regret_boost_factor=1.1,
               ewma_decay_factor=0.1):
    """Initializes the PolicyFromRegretAccumulator with parameters consistent with RegretAccumulator.

    Args:
      initial_optimism_factor: The initial weighting for the instantaneous regret component in the projection.
      optimism_decay_factor: Controls how quickly the optimism weight decays.
      positive_policy_exponent: Exponent for non-linear scaling of positive regrets.
      base_alpha: The baseline exponent for discounting positive regrets.
      base_beta: The baseline exponent for discounting negative regrets.
      volatility_sensitivity: Controls influence of volatility on both discounting
        and optimism dampening.
      max_expected_instantaneous_regret: Used for normalizing volatility.
      instantaneous_regret_boost_factor: Boost factor for positive instantaneous regrets.
      ewma_decay_factor: Decay factor for the EWMA of volatility.
    """
    self._initial_optimism_factor = initial_optimism_factor
    self._optimism_decay_factor = optimism_decay_factor
    self._positive_policy_exponent = positive_policy_exponent
    self._base_alpha = base_alpha
    self._base_beta = base_beta
    self._volatility_sensitivity = volatility_sensitivity
    self._max_expected_instantaneous_regret = max_expected_instantaneous_regret
    self._instantaneous_regret_boost_factor = instantaneous_regret_boost_factor
    self._ewma_decay_factor = ewma_decay_factor
    self._ewma_instantaneous_regret_magnitude = 0.0

  def get_updated_current_policy(self, info_state_node, iteration_number, cfr_regrets, previous_policy):
    """Obtains the current policy using a projection that is consistent with the RegretAccumulator's update rule.

    This method creates a tighter feedback loop by basing the current policy
    on a projection of what the regrets will be *after* the current
    iteration's update. This projection uses the same adaptive, asymmetric
    discounting and boosting logic as the main regret accumulation step.
    """
    # Centralize adaptive parameter calculation, ensuring consistency with RegretAccumulator.
    (
        self._ewma_instantaneous_regret_magnitude,
        normalized_volatility,
        discount_factor_positive,
        discount_factor_negative,
    ) = RegretAccumulator._calculate_adaptive_params(
        iteration_number=iteration_number,
        cfr_regrets=cfr_regrets,
        base_alpha=self._base_alpha,
        base_beta=self._base_beta,
        volatility_sensitivity=self._volatility_sensitivity,
        max_expected_instantaneous_regret=self._max_expected_instantaneous_regret,
        ewma_decay_factor=self._ewma_decay_factor,
        current_ewma_magnitude=self._ewma_instantaneous_regret_magnitude,
    )

    base_optimism = self._initial_optimism_factor / (1.0 + float(iteration_number) / self._optimism_decay_factor)
    # Dampen optimism during volatile periods to increase stability.
    optimism_dampening_factor = max(0.0, 1.0 - self._volatility_sensitivity * normalized_volatility)
    optimism_strength = base_optimism * optimism_dampening_factor

    action_to_projected_regret = {}
    for action in info_state_node.legal_actions:
      old_cumulative_regret = info_state_node.cumulative_regret.get(action, 0.0)
      instantaneous_regret = cfr_regrets.get(action, 0.0)

      instantaneous_regret_component = instantaneous_regret
      if instantaneous_regret_component > 0:
          instantaneous_regret_component *= self._instantaneous_regret_boost_factor

      if old_cumulative_regret >= 0:
          discounted_old_regret = discount_factor_positive * old_cumulative_regret
      else:
          discounted_old_regret = discount_factor_negative * old_cumulative_regret

      projected_regret = discounted_old_regret + optimism_strength * instantaneous_regret_component
      action_to_projected_regret[action] = projected_regret

    positive_scaled_projected_regrets = {
        action: (max(0.0, regret) ** self._positive_policy_exponent)
        for action, regret in action_to_projected_regret.items()
    }

    sum_positive_scaled_projected_regrets = sum(positive_scaled_projected_regrets.values())

    info_state_policy = {}
    if sum_positive_scaled_projected_regrets > 0:
      for action, scaled_regret in positive_scaled_projected_regrets.items():
        info_state_policy[action] = scaled_regret / sum_positive_scaled_projected_regrets
    else:
      num_legal_actions = len(info_state_node.legal_actions)
      for action in info_state_node.legal_actions:
        info_state_policy[action] = 1.0 / num_legal_actions

    return info_state_policy

class PolicyAccumulator:
  """A class that updates cumulative policy using regret-informed weighted
  averaging with a warmup period."""

  def __init__(self, base_gamma=2.0, gamma_max=4.0,
               gamma_volatility_sensitivity=1.5, warmup_iterations=500,
               stability_exponent=1.5, max_expected_instantaneous_regret=2.0,
               regret_magnitude_weighting_exponent=0.5): # New parameter
    """Initializes the PolicyAccumulator with adaptive gamma parameters and regret-magnitude weighting.

    Args:
      base_gamma: The baseline exponent for polynomial weighting of policies.
      gamma_max: The maximum value the adaptive gamma can reach.
      gamma_volatility_sensitivity: Controls how strongly volatility influences gamma.
      warmup_iterations: Number of initial iterations to skip for policy averaging.
      stability_exponent: Exponent for the stability factor based on regret magnitude.
      max_expected_instantaneous_regret: Normalization factor for instantaneous regret magnitude.
      regret_magnitude_weighting_exponent: Exponent for up-weighting policies based on
        the absolute magnitude of instantaneous regrets. Higher values give more
        emphasis to policies from iterations with large regrets.
    """
    self._base_gamma = base_gamma
    self._gamma_max = gamma_max
    self._gamma_volatility_sensitivity = gamma_volatility_sensitivity
    self._warmup_iterations = warmup_iterations
    self._stability_exponent = stability_exponent
    self._max_expected_instantaneous_regret = max_expected_instantaneous_regret
    self._regret_magnitude_weighting_exponent = regret_magnitude_weighting_exponent # Stored

  def update_accumulate_policy(
      self,
      info_state_node,
      iteration_number,
      info_state_policy,
      cfr_regrets,
      reach_prob,
      counterfactual_reach_prob,
  ):
    """Updates cumulative policy using delayed, regret-informed, and regret-magnitude weighted averaging.
    """
    if iteration_number < self._warmup_iterations:
      return info_state_node.cumulative_policy

    # Calculate instantaneous regret magnitude (L-infinity norm) for this iteration
    instantaneous_regret_magnitude = max(
        (abs(r) for r in cfr_regrets.values()), default=0.0
    )

    # Normalize volatility using the shared parameter
    if self._max_expected_instantaneous_regret > 0:
        normalized_volatility = min(1.0, instantaneous_regret_magnitude / self._max_expected_instantaneous_regret)
    else:
        normalized_volatility = 0.0

    # Adapt gamma based on volatility: higher volatility -> higher gamma
    effective_gamma = self._base_gamma + self._gamma_volatility_sensitivity * normalized_volatility
    effective_gamma = min(self._gamma_max, effective_gamma)

    # Standard polynomial weighting gives more weight to later iterations, now with adaptive gamma.
    temporal_weight = (float(iteration_number) + 1.0) ** effective_gamma

    # Calculate a stability factor from the L-infinity norm of instantaneous regrets.
    # Higher regret magnitude -> lower stability factor, using L-infinity norm for consistency.
    regret_stability_factor = 1.0 / (1.0 + instantaneous_regret_magnitude**self._stability_exponent)

    # NEW: Regret Magnitude Weighting Factor
    # Policies from iterations with higher regret magnitude contribute more to the average.
    # This factor boosts the weight, with higher values of the exponent giving more emphasis.
    # Ensure it's never zero to avoid division by zero or completely nullifying weight.
    regret_magnitude_factor = (
        1.0 + (instantaneous_regret_magnitude / self._max_expected_instantaneous_regret)
    ) ** self._regret_magnitude_weighting_exponent
    regret_magnitude_factor = max(0.1, regret_magnitude_factor) # Ensure minimum value to avoid zero weight if normalization results in very small number

    # The final weight combines the temporal, stability, and regret-magnitude-based components.
    weight = temporal_weight * regret_stability_factor * regret_magnitude_factor

    return {
        action: (
            info_state_node.cumulative_policy[action]
            + weight * reach_prob * info_state_policy[action]
        )
        for action in info_state_policy
    }
\end{lstlisting}

\begin{lstlisting}[
caption={SHOR-PSRO},
language=Python, 
label=lst:shor-psro,
]
import numpy as np

def _smoothed_best_pure_strategy(payoff_vec, temperature=1.0):
  """Computes a smoothed distribution biased towards the best pure strategy.
  The softmax function ensures that strategies with higher payoffs are
  given higher probability, with 'temperature' controlling the sharpness
  of the distribution. A lower temperature makes the distribution more
  concentrated on the best strategy, while a higher temperature
  leads to a more uniform distribution.
  """
  # Subtract max payoff for numerical stability (standard softmax trick)
  stable_payoffs = payoff_vec - np.max(payoff_vec)
  exp_payoffs = np.exp(stable_payoffs / temperature)
  
  sum_exp_payoffs = np.sum(exp_payoffs)
  if sum_exp_payoffs > 1e-12:  # Avoid division by zero
    return exp_payoffs / sum_exp_payoffs
  else:
    # Fallback to uniform distribution if all exponentiated payoffs are
    # effectively zero (e.g., due to very low temperature and negative payoffs,
    # or all payoffs being identical after stabilization).
    return np.ones_like(payoff_vec) / len(payoff_vec)

def _hybrid_orm_solver(meta_games, iterations,
                       blending_factor=0.0,
                       temperature=0.1,
                       momentum_beta=0.0,
                       gain_normalization=True,
                       diversity_bonus_coeff=0.0,
                       return_average_strategy=True): # New: Flag to return average or last-iterate strategy
  """Computes meta-strategies using Optimistic Regret Matching+ enhanced with
  optimistic updates, gain normalization, and a diversity bonus, then blended
  with a smoothed best pure strategy.

  This solver combines the stability and convergence properties of Optimistic
  Regret Matching+ (ORM+) with an explicit pull towards highly rewarding
  pure strategies, smoothed by a temperature-controlled softmax. This hybrid
  approach aims to leverage ORM+'s ability to find mixed equilibria while
  also quickly identifying and exploring strong pure-strategy modes in the
  meta-game, thereby potentially accelerating the discovery of low-exploitable
  policies in PSRO. The blending factor controls the trade-off between
  these two dynamics.

  Args:
    meta_games: A list of n-dimensional numpy arrays, one per player.
    iterations: Number of internal solver iterations.
    blending_factor: Weight (0 to 1) for blending ORM+ output with the
      smoothed best pure strategy. A factor of 0 means pure ORM+; 1 means
      pure smoothed best pure strategy.
    temperature: Temperature for softmax smoothing when calculating the
      smoothed best pure strategy. Lower values make the smoothing sharper.
    momentum_beta: Momentum parameter for optimistic updates to payoff gains.
    gain_normalization: If True, normalizes payoff gains to make learning rate
      more robust across games.
    diversity_bonus_coeff: Coefficient for diversity bonus, encouraging
      exploration of less-chosen policies.
    return_average_strategy: If True, returns time-averaged strategies.
      If False, returns last-iterate strategies.

  Returns:
    A list of mixed-strategies, one for each player, as numpy arrays.
  """
  num_players = len(meta_games)
  num_strats = [m.shape[i] for i, m in enumerate(meta_games)]

  if any(n_s == 0 for n_s in num_strats):
      return [np.array([]).tolist() for _ in range(num_players)]

  strategies = [np.ones(s, dtype=float) / s for s in num_strats]
  cum_regrets = [np.zeros(s, dtype=float) for s in num_strats]
  avg_strategies = [np.zeros(s, dtype=float) for s in num_strats]
  prev_centered_payoff_gains = [np.zeros(s, dtype=float) for s in num_strats]

  for t in range(iterations):
    current_centered_payoff_gains = [np.zeros(s, dtype=float) for s in num_strats]
    orm_strategies_this_iter = [np.zeros(s, dtype=float) for s in num_strats]

    for p in range(num_players):
      payoff_vec = meta_games[p]
      for other_p in reversed(range(num_players)):
        if other_p != p:
          payoff_vec = np.tensordot(payoff_vec, strategies[other_p], axes=([other_p], [0]))

      centered_payoff_gains = payoff_vec - np.mean(payoff_vec)
      current_centered_payoff_gains[p] = centered_payoff_gains

      optimistic_payoff_gains = (1 + momentum_beta) * centered_payoff_gains - \
                                 momentum_beta * prev_centered_payoff_gains[p]
      
      diversity_bonus = diversity_bonus_coeff * (1.0 - strategies[p])
      
      gains_for_regret_update = optimistic_payoff_gains + diversity_bonus
      
      if gain_normalization:
        max_abs_gain = np.max(np.abs(gains_for_regret_update))
        if max_abs_gain > 1e-8:
          gains_for_regret_update /= max_abs_gain

      cum_regrets[p] += gains_for_regret_update
      cum_regrets[p] = np.maximum(0, cum_regrets[p])

      sum_pos_regret = cum_regrets[p].sum()
      if sum_pos_regret > 1e-12:
        orm_strategies_this_iter[p] = cum_regrets[p] / sum_pos_regret
      else:
        orm_strategies_this_iter[p] = np.ones(num_strats[p]) / num_strats[p]
      
      smoothed_best_pure = _smoothed_best_pure_strategy(payoff_vec, temperature)
      
      strategies[p] = (1 - blending_factor) * orm_strategies_this_iter[p] + \
                      blending_factor * smoothed_best_pure

      prev_centered_payoff_gains[p] = current_centered_payoff_gains[p]

      if return_average_strategy: # Accumulate blended strategy only if average is requested
        avg_strategies[p] += strategies[p]

  if return_average_strategy:
    final_strategies = []
    for p in range(num_players):
      sum_avg_strat = np.sum(avg_strategies[p])
      if sum_avg_strat > 0:
        final_strategies.append(avg_strategies[p] / sum_avg_strat)
      else:
        final_strategies.append(np.ones(num_strats[p]) / num_strats[p])
    return final_strategies
  else:
    # If not returning average, return the last-iterate strategies
    return strategies

class TrainMetaStrategySolver:
  """A hybrid meta-solver for training that blends ORM+ with smoothed best pure strategies.

  This solver aims to accelerate convergence to low-exploitable strategies by
  dynamically balancing regret-minimization with a pull towards high-performing
  (but smoothed) pure strategies. Optimistic updates, gain normalization, and
  a diversity bonus are incorporated for improved learning dynamics. The
  blending factor, temperature, and diversity bonus are annealed over the
  outer PSRO iterations.
  """

  def __init__(self,
               base_solver_iterations=1000, # Base number of internal iterations
               iterations_per_policy_scale=20, # How much iterations scale per added policy
               max_solver_iterations=5000, # Max internal solver iterations
               initial_blending_factor=0.3, final_blending_factor=0.05,
               initial_temperature=0.5, final_temperature=0.01,
               momentum_beta=0.5,
               gain_normalization=True,
               initial_diversity_bonus_coeff=0.05,
               final_diversity_bonus_coeff=0.001,
               max_psro_iterations_for_annealing=75):
    """Initializes hybrid ORM solver parameters for training.

    Args:
      base_solver_iterations: Base number of internal solver iterations for _hybrid_orm_solver.
      iterations_per_policy_scale: Amount to increase internal solver iterations per added policy.
      max_solver_iterations: Maximum internal solver iterations.
      initial_blending_factor: Initial weight for the smoothed best pure strategy component.
      final_blending_factor: Final weight for the smoothed best pure strategy component.
      initial_temperature: Initial temperature for softmax smoothing.
      final_temperature: Final temperature for softmax smoothing.
      momentum_beta: Momentum for optimistic updates.
      gain_normalization: Normalizes gains for scale-invariance.
      initial_diversity_bonus_coeff: Max initial diversity bonus coefficient.
      final_diversity_bonus_coeff: Min initial diversity bonus coefficient across PSRO iterations.
      max_psro_iterations_for_annealing: PSRO iterations over which outer annealing occurs.
    """
    self._base_solver_iterations = base_solver_iterations
    self._iterations_per_policy_scale = iterations_per_policy_scale
    self._max_solver_iterations = max_solver_iterations
    self._initial_blending_factor = initial_blending_factor
    self._final_blending_factor = final_blending_factor
    self._initial_temperature = initial_temperature
    self._final_temperature = final_temperature
    self._momentum_beta = momentum_beta
    self._gain_normalization = gain_normalization
    self._initial_diversity_bonus_coeff = initial_diversity_bonus_coeff
    self._final_diversity_bonus_coeff = final_diversity_bonus_coeff
    self._max_psro_iterations_for_annealing = max_psro_iterations_for_annealing
    self._current_psro_iteration = 0

  def get_meta_strategy(self, game, policy_sets, meta_games):
    """Returns blended meta strategies for training.

    Args:
      game: The pyspiel game object.
      policy_sets: A list of lists of policies, one list per player.
        policy_sets[p][i] is player p's i-th policy. len(policy_sets[p]) ==
        meta_games[0].shape[p].
      meta_games: A list of n-dimensional numpy arrays, one per player. Each
        array has shape (num_strats_p0, num_strats_p1, ..., num_strats_pn-1) and
        meta_games[p][i0, i1, ..., in-1] is the payoff of player p when player k
        chooses strategy ik.

    Returns:
      A list of blended mixed-strategies.
    """
    del game, policy_sets # Unused

    self._current_psro_iteration += 1
    current_psro_iter = self._current_psro_iteration

    # Adaptive solver iterations: scale with current population size
    num_current_policies_p0 = len(meta_games[0]) # Assuming symmetric populations
    solver_iterations = int(self._base_solver_iterations +
                            self._iterations_per_policy_scale * (num_current_policies_p0 - 1))
    solver_iterations = np.clip(solver_iterations, self._base_solver_iterations, self._max_solver_iterations)

    annealing_progress = min(1.0, current_psro_iter / self._max_psro_iterations_for_annealing)

    blending_factor = (self._initial_blending_factor * (1.0 - annealing_progress) +
                       self._final_blending_factor * annealing_progress)
    
    temperature = (self._initial_temperature * (1.0 - annealing_progress) +
                   self._final_temperature * annealing_progress)

    diversity_bonus_coeff = (self._initial_diversity_bonus_coeff * (1.0 - annealing_progress) +
                             self._final_diversity_bonus_coeff * annealing_progress)
    
    blending_factor = np.clip(blending_factor, self._final_blending_factor, self._initial_blending_factor)
    temperature = np.clip(temperature, self._final_temperature, self._initial_temperature)
    diversity_bonus_coeff = np.clip(diversity_bonus_coeff, self._final_diversity_bonus_coeff, self._initial_diversity_bonus_coeff)

    strategies = _hybrid_orm_solver(
        meta_games,
        iterations=solver_iterations, # Use adaptive iterations
        blending_factor=blending_factor,
        temperature=temperature,
        momentum_beta=self._momentum_beta,
        gain_normalization=self._gain_normalization,
        diversity_bonus_coeff=diversity_bonus_coeff,
        return_average_strategy=True # Training always uses averaged strategies for stability
    )
    return [s.tolist() for s in strategies]

class EvalMetaStrategySolver:
  """Returns meta strategies for evaluation in PSRO.

  This solver uses a hybrid approach, blending Optimistic Regret Matching+
  with a smoothed best pure strategy, tailored for robust and accurate
  exploitability measurement. The parameters are set to emphasize
  exploitation for evaluation purposes, including optimistic updates and
  gain normalization for stability, while keeping diversity bonus minimal.
  Crucially, it returns the *last-iterate* strategy for a reactive estimate
  of exploitability.
  """

  def __init__(self,
               base_solver_iterations=8000, # Base number of internal iterations
               iterations_per_policy_scale=50, # How much iterations scale per added policy
               max_solver_iterations=15000, # Max internal solver iterations
               blending_factor=0.01,
               temperature=0.001,
               momentum_beta=0.2,
               gain_normalization=True,
               diversity_bonus_coeff=0.0):
    """Initializes hybrid ORM solver parameters for evaluation meta-strategies.

    Args:
      base_solver_iterations: Base number of internal solver iterations for _hybrid_orm_solver.
      iterations_per_policy_scale: Amount to increase internal solver iterations per added policy.
      max_solver_iterations: Maximum internal solver iterations.
      blending_factor: Weight (0 to 1) for the smoothed best pure strategy component.
      temperature: Temperature for softmax smoothing.
      momentum_beta: Momentum for optimistic updates.
      gain_normalization: Normalizes gains for scale-invariance.
      diversity_bonus_coeff: Diversity bonus, kept very low for evaluation.
    """
    self._base_solver_iterations = base_solver_iterations
    self._iterations_per_policy_scale = iterations_per_policy_scale
    self._max_solver_iterations = max_solver_iterations
    self._blending_factor = blending_factor
    self._temperature = temperature
    self._momentum_beta = momentum_beta
    self._gain_normalization = gain_normalization
    self._diversity_bonus_coeff = diversity_bonus_coeff

  def get_meta_strategy(self, game, policy_sets, meta_games):
    """Returns blended meta strategies for evaluation in policy-space response oracles.

    Args:
      game: The pyspiel game object.
      policy_sets: A list of lists of policies, one list per player.
        policy_sets[p][i] is player p's i-th policy. len(policy_sets[p]) ==
        meta_games[0].shape[p].
      meta_games: A list of n-dimensional numpy arrays, one per player. Each
        array has shape (num_strats_p0, num_strats_p1, ..., num_strats_pn-1) and
        meta_games[p][i0, i1, ..., in-1] is the payoff of player p when player k
        chooses strategy ik.

    Returns:
      A list of mixed-strategies, one for each player. Each mixed strategy is
      a list of non-negative weights (not necessarily normalized). It is used
      for evaluation of the current PSRO policies. E.g., computing
      exploitability.
    """
    del game, policy_sets # Unused

    num_current_policies_p0 = len(meta_games[0]) # Assuming symmetric populations
    solver_iterations = int(self._base_solver_iterations +
                            self._iterations_per_policy_scale * (num_current_policies_p0 - 1))
    solver_iterations = np.clip(solver_iterations, self._base_solver_iterations, self._max_solver_iterations)

    strategies = _hybrid_orm_solver(
        meta_games,
        iterations=solver_iterations, # Use adaptive iterations
        blending_factor=self._blending_factor,
        temperature=self._temperature,
        momentum_beta=self._momentum_beta,
        gain_normalization=self._gain_normalization,
        diversity_bonus_coeff=self._diversity_bonus_coeff,
        return_average_strategy=False # Eval explicitly requests last-iterate strategy
    )
    return [s.tolist() for s in strategies]
\end{lstlisting}

\subsubsection{Detailed Specification of VAD-CFR Ablation Components}
\label{sec:vad_ablations}

To understand the contribution of each mechanism discovered by AlphaEvolve within the Volatility-Adaptive Discounted (VAD-) CFR, we conducted a systematic ablation study. Each ablation removes or reverts a specific evolved component to its standard Counterfactual Regret Minimization (CFR) or Discounted CFR (DCFR) equivalent. The following variants correspond to the empirical results presented in Table~\ref{tab:vad_cfr_audit}.

\paragraph{Regret Accumulation Variants}
\begin{itemize}
    
    \item \textbf{VAD-CFR (no\_asym\_discount):} VAD-CFR utilizes highly asymmetric discounting factors for positive and negative regrets. This ablation forces a symmetric baseline discount factor by setting \texttt{base\_beta = 1.5} (matching \texttt{base\_alpha}), testing the hypothesis that an aggressive ``forget-gate'' ($\beta = -0.1$) is necessary for recovery in non-stationary multi-agent environments.
    
    \item \textbf{VAD-CFR (no\_boost):} This variant removes the ``Exploration Boost'' mechanism by reverting \texttt{instantaneous\_regret\_boost\_factor} to $1.0$. In the full algorithm, the $1.1$ boost accelerates the accumulation of regrets for under-explored, high-potential actions.
\end{itemize}

\paragraph{Policy Accumulation and Strategy Distillation}
\begin{itemize}
    \item \textbf{VAD-CFR (no\_predication):} VAD-CFR uses a non-linear projection when deriving the current iteration policy from accumulated regrets. This ablation reverts the \texttt{PolicyFromRegretAccumulator} to a standard linear Regret Matching mapping by setting \texttt{positive\_policy\_exponent = 1.0} and removing the anticipatory \texttt{initial\_optimism\_factor = 0.0}.
    
    \item \textbf{VAD-CFR (no\_warmstart):} One of the most significant discovered features is a hard warm-start delay. This ablation bypasses the \texttt{warmup\_iterations = 500} condition, forcing the algorithm to begin policy averaging from $t=1$. This tests whether a ``burn-in'' period is critical for preventing early, high-variance policies from polluting the final average strategy.
\end{itemize}

\paragraph{Environmental Adaptation}
\begin{itemize}
    \item \textbf{VAD-CFR (no\_volatility):} The core dynamic of VAD-CFR is its ability to adjust discounting parameters based on the perceived volatility of the game state, tracked via an Exponential Weighted Moving Average (EWMA) of instantaneous regrets. This ablation disables the \texttt{calculate\_adaptive\_params} logic, replacing the adaptive volatility scaling with static hyperparameters. This directly tests the value of the meta-adaptation dynamics discovered by the LLM versus a fixed discount schedule.
\end{itemize}

\subsubsection{Code of WOP-CFR}
See Listing~\ref{lst:wop_cfr}.

\begin{figure}
\begin{lstlisting}[language=Python, caption={WOP-CFR}, label={lst:wop_cfr}]
class WOPRegretAccumulator:
  def update_accumulate_regret(self, node, iter_num, cfr_regrets):
    t = float(iter_num + 1)
    d_pos, d_neg = (t**1.5)/(t**1.5 + 1.0), (t**-0.1)/(t**-0.1 + 1.0)
    return {a: (d_pos if node.cumulative_regret[a] >= 0 else d_neg) * node.cumulative_regret[a] + cfr_regrets[a] for a in cfr_regrets}

class WOPPolicyFromRegretAccumulator:
  def get_updated_current_policy(self, node, iter_num, cfr_regrets, previous_policy):
    t = float(iter_num + 1)
    d_pos, d_neg = (t**1.5)/(t**1.5 + 1.0), (t**-0.1)/(t**-0.1 + 1.0)
    opt_s = 1.0 / (1.0 + float(iter_num) / 100.0)
    proj = {}
    for a in node.legal_actions:
      d = d_pos if node.cumulative_regret.get(a, 0.0) >= 0 else d_neg
      r = d * node.cumulative_regret.get(a, 0.0) + opt_s * cfr_regrets.get(a, 0.0)
      proj[a] = max(0.0, r) ** 1.5
    total = sum(proj.values())
    return {a: v/total if total > 0 else 1.0/len(proj) for a, v in proj.items()}

class WOPPolicyAccumulator:
  def update_accumulate_policy(self, node, iter_num, policy, cfr_regrets, reach, cf_reach):
    if iter_num < 500: return node.cumulative_policy # Hard Warm-Start
    weight = (float(iter_num) + 1.0) ** 2.0 # Static Polynomial Weight
    return {a: node.cumulative_policy[a] + weight * reach * policy[a] for a in policy}
\end{lstlisting}
\end{figure}

\subsubsection{Detailed Specification of SHOR-PSRO Ablation Components}
\label{sec:psro_ablations}

To understand the contribution of each mechanism discovered by AlphaEvolve within the Smoothed Hybrid Optimistic Regret (SHOR-) PSRO framework, we conducted a comprehensive ablation study. The raw SHOR-PSRO algorithm relies on a highly parameterized, asymmetric design. Each ablation isolates a specific heuristic by either neutralizing its parameters or reverting the logic to standard baselines. The following variants correspond to the empirical results presented in Section~\ref{sec:exp_shorpsro}, with direct references to the algorithmic components defined in Listing~\ref{lst:shor-psro}.

\textbf{Train/Eval Asymmetry Ablations} \\
The raw LLM discovery evolved two distinct, highly specialized solvers for the training-time Oracle generation and the evaluation-time exploitability measurement.
\begin{itemize}
    \item \textbf{SHOR-PSRO (train\_as\_eval):} This ablation forces the evaluation phase to utilize the \texttt{SHORTrainMetaStrategySolver}. It tests whether the exploratory, high-temperature heuristics used to generate new policies are fundamentally detrimental when used to measure exact exploitability.
    \item \textbf{SHOR-PSRO (eval\_as\_train):} This ablation forces the training phase to utilize the \texttt{SHOREvalMetaStrategySolver}, stripping away the dynamic annealing and diversity bonuses during population expansion to test if the training solver's complexity is genuinely necessary for exploration.
\end{itemize}

\textbf{Blending and Annealing Ablations} \\
SHOR-PSRO's core innovation is the active interpolation between an Optimistic Regret Matching+ (ORM+) loop and a smoothed best pure strategy, guided by automated schedules.
\begin{itemize}
    \item \textbf{SHOR-PSRO (no\_blending\_pure):} Removes the ORM+ engine entirely by strictly setting \texttt{initial\_blending\_factor=1.0} and \texttt{final\_blending\_factor=1.0}. The solver relies exclusively on the temperature-smoothed pure strategy.
    \item \textbf{SHOR-PSRO (no\_blending\_orm):} Removes the pure strategy bias by strictly setting \texttt{initial\_blending\_factor=0.0} and \texttt{final\_blending\_factor=0.0}. The solver relies exclusively on the ORM+ update loop.
    \item \textbf{SHOR-PSRO (no\_annealing):} Disables the dynamic scaling of hyperparameters across PSRO epochs. It freezes the \texttt{blending\_factor} at $0.05$, \texttt{temperature} at $0.01$, and \texttt{diversity\_bonus\_coeff} at $0.001$, preventing the solver from transitioning from exploration to exploitation as the population grows.
    \item \textbf{SHOR-PSRO (fixed\_iterations):} Disables the \texttt{iterations\_per\_policy\_scale} mechanism, testing whether scaling the internal solver iterations dynamically with the size of the empirical game tensor is a critical mechanism or an over-parameterized heuristic.
\end{itemize}

\textbf{ORM Engine Heuristic Ablations} \\
The underlying ORM+ loop discovered by the LLM contains several specialized micro-mechanics intended to accelerate convergence.
\begin{itemize}
    \item \textbf{SHOR-PSRO (no\_diversity):} Removes the uniform exploration incentive by strictly setting \texttt{diversity\_bonus\_coeff=0.0}.
    \item \textbf{SHOR-PSRO (no\_momentum):} Disables the optimistic update of the payoff gains by setting \texttt{momentum\_beta=0.0}, reverting the update rule closer to standard Regret Matching.
    \item \textbf{SHOR-PSRO (no\_gain\_norm):} Sets \texttt{gain\_normalization=False}, disabling the adaptive scaling of the optimistic gains matrix, which tests the algorithm's vulnerability to extreme payoff magnitudes.
\end{itemize}

\begin{table}[ht]
\centering
\caption{Exhaustive Algorithmic Audit of SHOR-PSRO Components. Results are reported as IQM Log-Improvement Scores ($s_g$).}
\label{tab:exhaustive_psro_ablations}
\begin{tabular}{lccc}
\toprule
\textbf{Variant} & \textbf{Total IQM} & \textbf{Train IQM} & \textbf{Test IQM} \\
\midrule
\multicolumn{4}{l}{\textit{Distilled Solver}} \\
PM-PSRO & \textbf{+0.032} & \textbf{-0.038} & \textbf{+0.059} \\
\midrule
\multicolumn{4}{l}{\textit{Train/Eval Asymmetry Ablations}} \\
train\_as\_eval & +0.048 & +0.224 & +0.021 \\
eval\_as\_train & -0.557 & -0.584 & -0.541 \\
\midrule
\multicolumn{4}{l}{\textit{Blending \& Annealing Ablations}} \\
no\_blending\_pure (Removes ORM entirely) & \textbf{-1.525} & -\textbf{2.292} & \textbf{-1.268} \\
no\_blending\_orm (Removes pure blend) & +0.012 & -0.069 & +0.020 \\
no\_annealing & +0.013 & -0.104 & +0.016 \\
fixed\_iterations & -0.038 & -0.070 & -0.031 \\
\midrule
\multicolumn{4}{l}{\textit{ORM Engine Heuristic Ablations}} \\
no\_diversity & +0.007 & -0.015 & +0.012 \\
no\_momentum & -0.017 & -0.018 & -0.016 \\
no\_gain\_norm & -0.012 & +0.012 & -0.017 \\
\bottomrule
\end{tabular}
\end{table}

\subsubsection{Code of PM-PSRO}
See Listing~\ref{lst:pgm_plus_psro}.
\begin{figure}
\begin{lstlisting}[language=Python, caption={PM-PSRO.}, label={lst:pgm_plus_psro}]
import numpy as np

class PMMetaStrategySolver:
    """Projection Matching (PM) PSRO Solver.
    A minimal meta-solver utilizing geometric tangent 
    projection, and a dynamic iteration budget.
    """
    def __init__(self, base_iterations=1000, iter_scale=20):
        self.base_iterations = base_iterations
        self.iter_scale = iter_scale

    def get_meta_strategy(self, game, policy_sets, meta_games):
        num_players = len(meta_games)
        num_strats = [m.shape[i] for i, m in enumerate(meta_games)]
        num_current_policies = num_strats[0]
        iterations = self.base_iterations + self.iter_scale * (num_current_policies - 1)
        strategies = [np.ones(s) / s for s in num_strats]
        duals = [np.zeros(s) for s in num_strats]
        avg_strategies = [np.zeros(s) for s in num_strats]
        for _ in range(iterations):
            for p in range(num_players):
                payoff_vec = meta_games[p]
                for other_p in reversed(range(num_players)):
                    if other_p != p:
                        payoff_vec = np.tensordot(
                            payoff_vec, strategies[other_p], axes=([other_p], [0])
                        )

                projected_grad = payoff_vec - np.mean(payoff_vec)
                duals[p] += projected_grad
                positive_duals = np.maximum(duals[p], 0)
                sum_duals = np.sum(positive_duals)
                if sum_duals > 0:
                    strategies[p] = positive_duals / sum_duals
                else:
                    strategies[p] = np.ones(num_strats[p]) / num_strats[p]
                avg_strategies[p] += strategies[p]

        final_strategies = []
        for avg, s in zip(avg_strategies, num_strats):
            sum_avg = np.sum(avg)
            if sum_avg > 0:
                final_strategies.append((avg / sum_avg).tolist())
            else:
                final_strategies.append((np.ones(s) / s).tolist())     
        return final_strategies
\end{lstlisting}
\end{figure}

  \subsection{Evaluated Game Suite}
  \label{app:games}
  
   We evaluate all algorithms on a suite of 18 game instances from the OpenSpiel framework~\citep{lanctot2019openspiel}, spanning five game families. The suite is designed to cover a wide range of structural properties: sequential vs.\ simultaneous moves, 2--4 players, and game trees ranging from 2 to 347{,}810 information states with branching factors from 1 to 66. All games in our suite involve imperfect information. Table~\ref{tab:game_suite} provides a summary; detailed descriptions follow.
  
  \begin{table}[h]
  \centering
  \caption{Summary of the 18 evaluated game instances. All games involve imperfect information. ``Seq'' = sequential, ``Sim'' = simultaneous (converted to sequential via \texttt{turn\_based\_simultaneous\_game}). Information state counts and branching factors are exact, computed by full game tree traversal.}
  \label{tab:game_suite}
  \small
  \begin{tabular}{llccrcc}
  \toprule
  \textbf{Game Instance} & \textbf{Family} & \textbf{Players} & \textbf{Type} & \textbf{Info States} & \textbf{Avg.\ Actions} & \textbf{Action Range} \\
  \midrule
  Kuhn Poker & Poker & 2 & Seq & 12 & 2.0 & 2 \\
  Kuhn Poker (3p) & Poker & 3 & Seq & 48 & 2.0 & 2 \\
  Kuhn Poker (4p) & Poker & 4 & Seq & 160 & 2.0 & 2 \\
  Leduc Poker & Poker & 2 & Seq & 936 & 2.3 & 2--3 \\
  Leduc Poker (3p) & Poker & 3 & Seq & 25{,}800 & 2.3 & 2--3 \\
  Universal Poker & Poker & 2 & Seq & 20{,}160 & 2.4 & 2--4 \\
  \midrule
  Goofspiel-3 & Goofspiel & 2 & Sim & 114 & 2.1 & 2--3 \\
  Goofspiel-4 & Goofspiel & 2 & Sim & 6{,}056 & 2.1 & 2--4 \\
  Goofspiel-5 & Goofspiel & 2 & Sim & 347{,}810 & 2.1 & 2--5 \\
  Goofspiel-3 (limited info) & Goofspiel & 2 & Sim & 90 & 2.1 & 2--3 \\
  Goofspiel-4 (limited info) & Goofspiel & 2 & Sim & 3{,}608 & 2.1 & 2--4 \\
  Goofspiel-5 (limited info) & Goofspiel & 2 & Sim & 236{,}450 & 2.1 & 2--5 \\
  \midrule
  Liar's Dice (4-sided) & Liar's Dice & 2 & Seq & 1{,}024 & 2.0 & 1--8 \\
  Liar's Dice (5-sided) & Liar's Dice & 2 & Seq & 5{,}120 & 2.0 & 1--10 \\
  Liar's Dice (6-sided) & Liar's Dice & 2 & Seq & 24{,}576 & 2.0 & 1--12 \\
  \midrule
  Colonel Blotto & Blotto & 2 & Sim & 2 & 66.0 & 66 \\
  \midrule
  Battleship ($2{\times}2$) & Battleship & 2 & Seq & 10{,}198 & 4.0 & 4 \\
  Battleship ($3{\times}2$) & Battleship & 2 & Seq & 160{,}491 & 6.0 & 6--7 \\
  \bottomrule
  \end{tabular}
  \end{table}
  
  \subsubsection{Poker Variants}
  
  \paragraph{Kuhn Poker~\citep{kuhn1950simplified}.}
  A simplified poker game played with a three-card deck (Jack, Queen, King). Each player is dealt one card and places a single ante. Players then engage in a single round of betting with check/bet/call/fold actions. Despite its simplicity, Kuhn Poker has a non-trivial mixed Nash equilibrium and serves as a canonical benchmark for imperfect-information game-solving algorithms. We evaluate on the standard 2-player version (12 information states) and extended variants with 3 players (48 information states, 4-card deck) and 4 players (160 information states, 5-card deck).
  
  \paragraph{Leduc Poker~\citep{southey2005bayes}.}
  A two-round poker game using a deck of six cards (two suits $\times$ three ranks). Each player is dealt a single private card, followed by a round of betting; then a single community card is dealt, followed by a second betting round. Hands are ranked by pair (private card matches community card) and then by card rank. Leduc Poker is a standard benchmark for imperfect-information algorithms as it is small enough for exact Nash equilibrium computation yet complex enough to test key algorithmic properties. We evaluate on the 2-player version (936 information states) and a 3-player variant (25{,}800 information states).
  
  \paragraph{Universal Poker.}
  OpenSpiel's parameterized poker game that can represent arbitrary limit and no-limit hold'em variants through the ACPC (Annual Computer Poker Competition) protocol. In its default configuration, it implements a small limit hold'em game with a 4-card deck, a single betting round, and fixed bet sizes. With 20{,}160 information states, it provides a more complex poker benchmark than Leduc while remaining computationally tractable for tabular methods.
  
  \subsubsection{Goofspiel (The Game of Pure Strategy)}
  
  \paragraph{Goofspiel~\citep{ross1971goofspiel}.}
  A simultaneous-move card game in which two players compete to win ``prize'' cards using their ``bid'' cards. Each player holds a hand of cards numbered 1 through $n$, and a sequence of prize cards (also 1 through $n$) is revealed one at a time. In each round, both players simultaneously play a bid card; the higher bid wins the current prize (ties result in the prize being discarded). The player who accumulates the highest total prize value wins. Since players choose actions simultaneously, the game inherently involves imperfect information---each player must commit to a bid without knowing the opponent's choice. We convert the simultaneous-move game to sequential form via the \texttt{turn\_based\_simultaneous\_game} wrapper for compatibility with CFR-based algorithms.
  
  We evaluate on six variants organized along two dimensions:
  \begin{itemize}
      \item \textbf{Scale:} $n \in \{3, 4, 5\}$ cards, yielding progressively larger game trees (114 to 347{,}810 information states).
      \item \textbf{Information structure:} In the standard version, both players observe each other's bids after each round. In the \emph{limited information} version (\texttt{imp\_info=True}), players only observe whether they won or lost each round, but not the opponent's specific bid. Interestingly, the limited-information variant has \emph{fewer} information states (e.g., 236{,}450 vs.\ 347{,}810 for $n{=}5$) because more game histories are observationally equivalent from each player's perspective.
  \end{itemize}
  
  \subsubsection{Liar's Dice}
  
  \paragraph{Liar's Dice~\citep{lisy2015online}.}
  A bluffing game in which each player secretly rolls a single die, then players alternate making increasingly higher claims about the total count of a specific face value across \emph{all} dice. A player may either raise the current claim or challenge (``call liar'') the previous player's claim. If challenged, all dice are revealed: the challenger wins if the claim was false, and the claimant wins if it was true. We evaluate on variants with $d \in \{4, 5, 6\}$ die faces, yielding 1{,}024, 5{,}120, and 24{,}576 information states respectively.
  
  \subsubsection{Colonel Blotto}
  
  \paragraph{Colonel Blotto~\citep{borel1953theory}.}
  A classic resource-allocation game in which two players simultaneously distribute a fixed budget of units across multiple battlefields. Each battlefield is won by the player who allocates more units to it (ties broken randomly). The player who wins the most battlefields wins the game. In the default OpenSpiel configuration, each player has 10 units to distribute across 3 battlefields (66 possible allocations per player). As a single simultaneous decision, the game has only 2 information states (one per player) but a rich $66 \times 66$ payoff matrix with complex mixed Nash equilibria. The game is converted to sequential form via \texttt{turn\_based\_simultaneous\_game}.
  
  \subsubsection{Battleship}
  
  \paragraph{Battleship~\citep{farina2019correlation}.}
  A parametric two-player game inspired by the classic board game, formalized for game-theoretic analysis. Each player secretly places ships on a grid, then players alternate taking shots to locate and sink the opponent's ships. The game has imperfect information (players do not observe the opponent's ship placement) and is parameterized by board dimensions, ship configurations, and scoring rules. We evaluate on two configurations:
  \begin{itemize}
      \item \textbf{Battleship $2{\times}2$:} A $2{\times}2$ board with one ship of size 2 and value 2.0, with 3 shots allowed (10{,}198 information states).
      \item \textbf{Battleship $3{\times}2$:} A $3{\times}2$ board with the same ship and shot configuration (160{,}491 information states).
  \end{itemize}
  
  \subsubsection{Game Selection Rationale}
  
  The suite is designed to stress-test algorithm performance across several axes:
  \begin{enumerate}
      \item \textbf{Number of players:} Two-player (most games) vs.\ multi-player (Kuhn 3p/4p, Leduc 3p).
      \item \textbf{Game tree scale:} From 2 information states (Blotto) to 347{,}810 (Goofspiel-5), testing scalability.
      \item \textbf{Strategic depth:} From pure bluffing (poker, Liar's Dice) to resource allocation (Blotto) to bidding under uncertainty (Goofspiel).
      \item \textbf{Move structure:} Sequential (poker, Liar's Dice, Battleship) vs.\ simultaneous converted to sequential (Goofspiel, Blotto).
      \item \textbf{Observability:} Varying degrees of information---from fully observable bids (Goofspiel) to hidden bids with outcome feedback only (Goofspiel limited info) to hidden card placement (Battleship).
  \end{enumerate}

\subsection{Results on 18 Games}\label{sec:full_results}
\begin{figure*}[ht]
    \centering
    \includegraphics[width=1\textwidth]{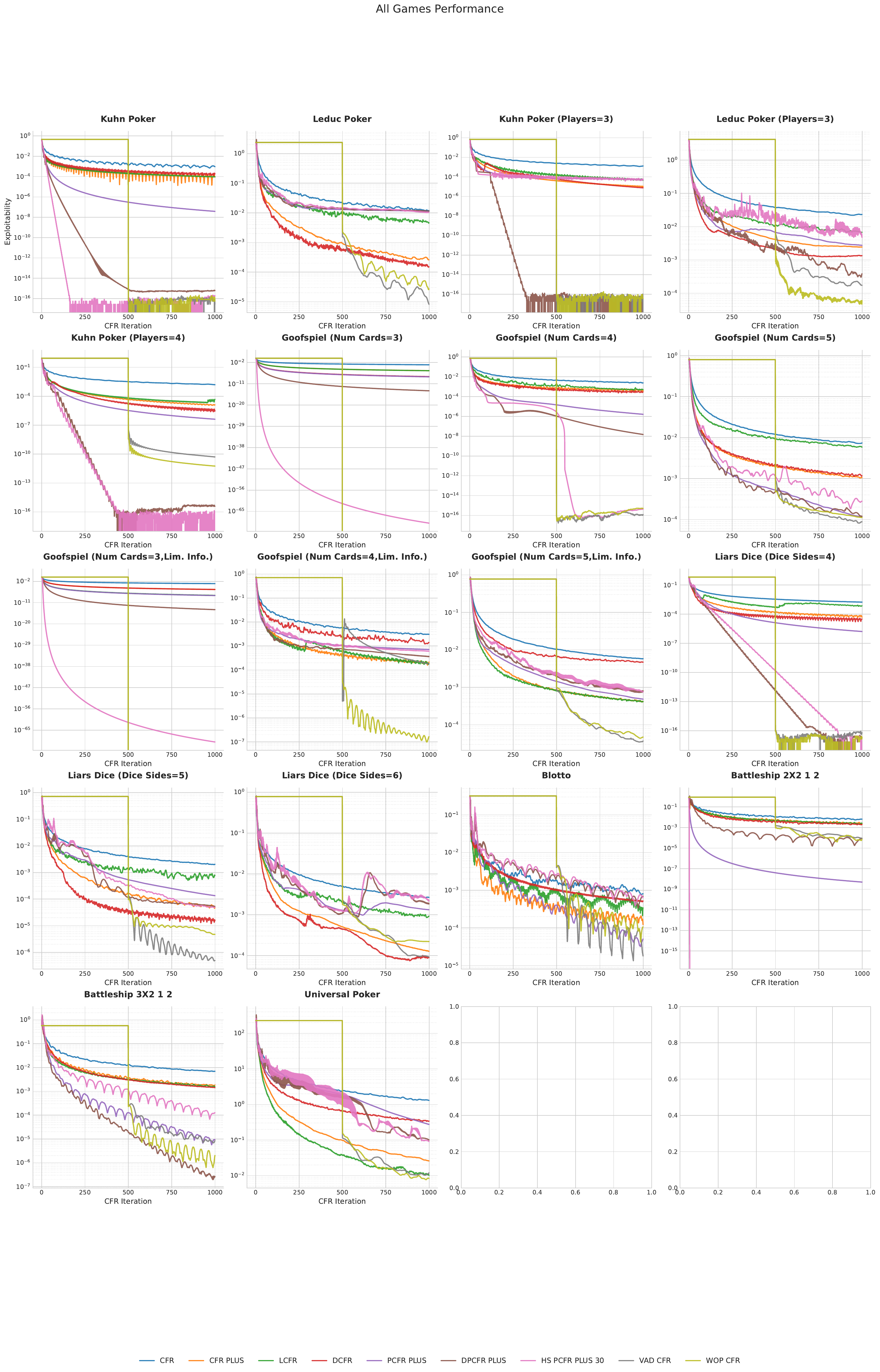}
    \caption{\textbf{CFR variants performances on All Games.}}
    \label{fig:cfr_all_games}
\end{figure*}

\begin{figure*}[ht]
    \centering
    \includegraphics[width=1\textwidth]{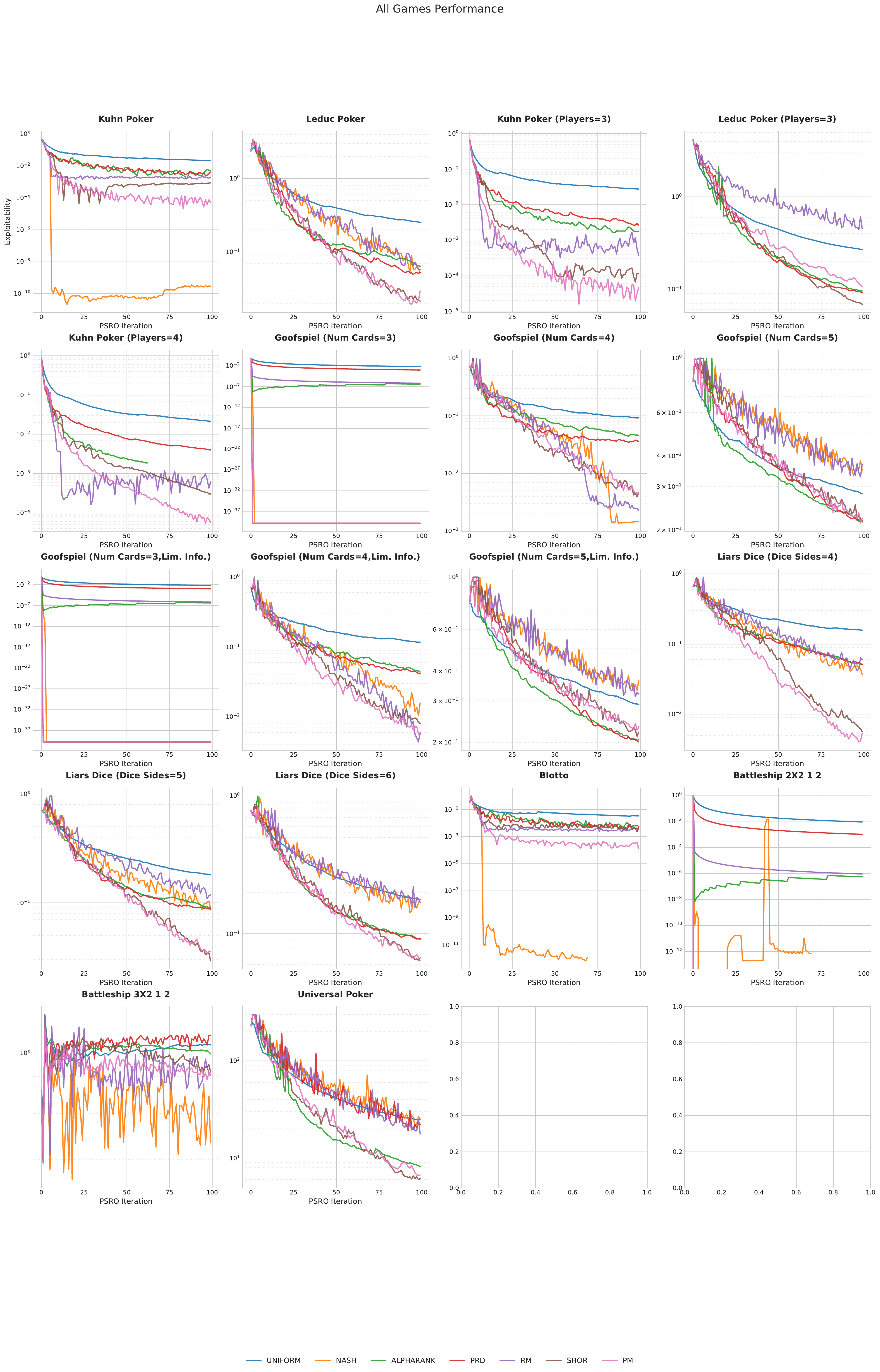}
    \caption{\textbf{PSRO variants performances on All Games.}}
    \label{fig:psro_all_games}
\end{figure*}

Figures~\ref{fig:cfr_all_games} and \ref{fig:psro_all_games} show full per-game exploitability curves for all 18 games in our evaluation suite for CFR and PSRO variants respectively.

\subsection{Additional PSRO Evaluation on Randomly Generated Normal-Form Games}\label{sec:nfg_results}
To complement the OpenSpiel benchmark with an evaluation that is structurally independent of any specific extensive-form game, we evaluate all PSRO meta-solvers on randomly generated $N$-player normal-form games (NFGs). This setting tests the meta-solver itself in isolation, removing any influence of the underlying extensive-form structure that AlphaEvolve might have implicitly overfit to during training. 
Sampling 1000 NFGs per configuration also provides a large-sample statistical complement to the deterministic per-game evaluation in the main body.

\paragraph{Game generation.} For each configuration $(N, A)$, we sample 1000 random $N$-player NFGs with $A$ actions per player. 
Each game is constructed by drawing independent payoff tensors $u_i \in \mathbb{R}^{A^N}$ for each player from either $\mathcal{N}(0, 1)$ (Gaussian) or $\mathcal{U}[-1, 1]$ (Uniform), and then centering across players to enforce constant-sum structure:

$$
u_i \leftarrow u_i - \tfrac{1}{N} \sum\nolimits_{j=1}^{N} u_j.
$$

After centering, $\sum_i u_i(a) = 0$ for every joint action profile $a$, generalizing two-player zero-sum to the $N$-player setting. Distinct random seeds are used across the 1000 sampled games and the two distributions.

\paragraph{Why $N \geq 3$.} We restrict the NFG evaluation to $N \geq 3$ because two-player zero-sum NFGs admit a polynomial-time exact solution via linear programming. In that regime the LP-Nash baseline trivially dominates all other meta-solvers and the comparison contains no signal about meta-solver design choices.

\paragraph{Experimental setup.} Each PSRO run starts from a single random pure strategy per player, expands the population by one exact best response per iteration, and terminates after $K = A$ outer iterations (full population growth). We benchmark Uniform, AlphaRank, PRD, RM, SHOR-PSRO, and PM-PSRO under identical conditions; the LP-Nash baseline is omitted as it does not extend to $N \geq 3$. Following~\citet{agarwal2021deep}, we report the per-iteration interquartile mean (IQM) of exploitability across the 1000 sampled games, with 95\% confidence intervals computed by 50,000 bootstrap resamples at the run level (i.e., resampling games rather than individual iterations, which preserves within-trajectory correlation).

\begin{figure*}[ht]
    \centering
    \includegraphics[width=1\textwidth]{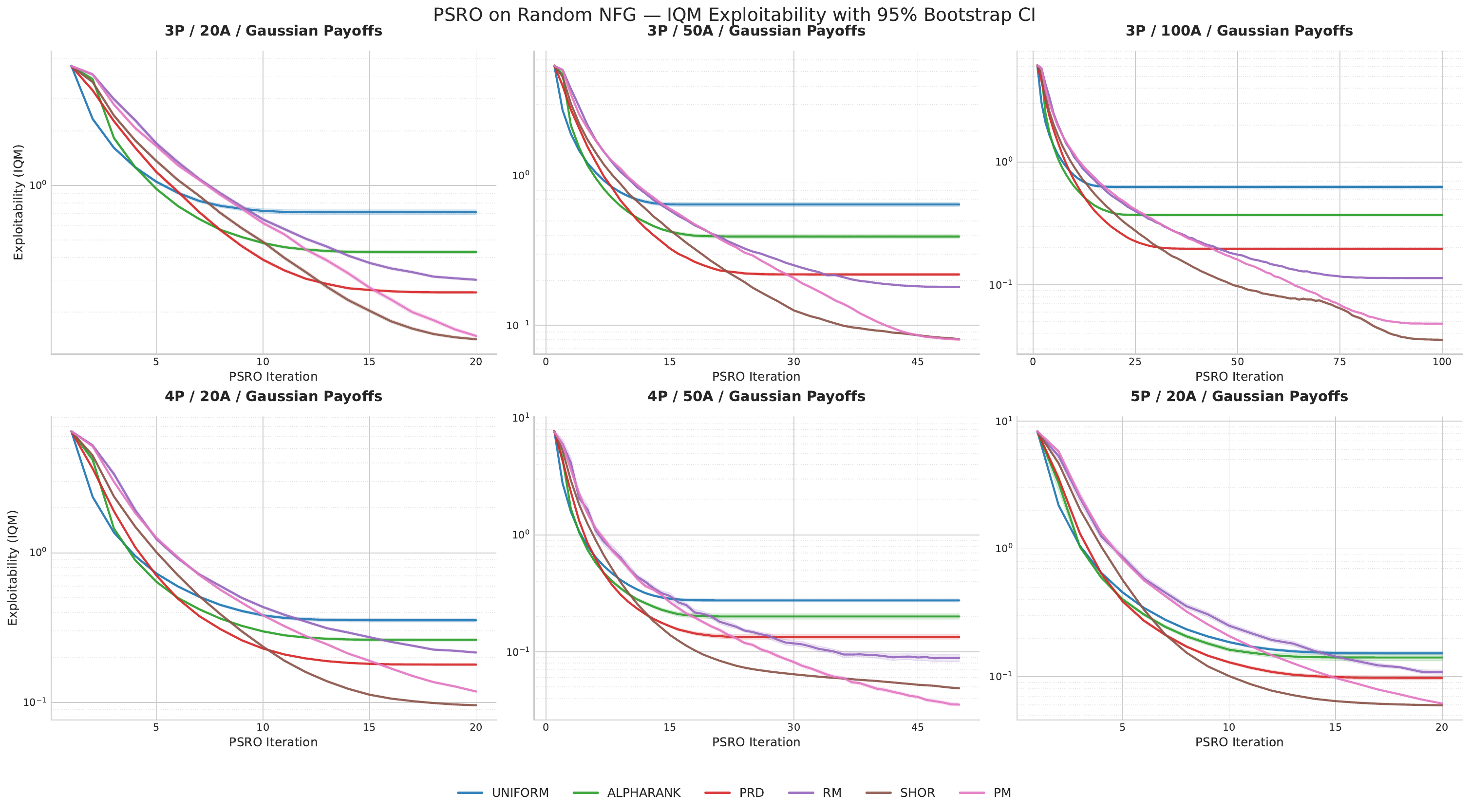}
    \caption{\textbf{PSRO variants performances on Gaussian payoff games.}}
    \label{fig:gaussian_nfg}
\end{figure*}

\begin{figure*}[ht]
    \centering
    \includegraphics[width=1\textwidth]{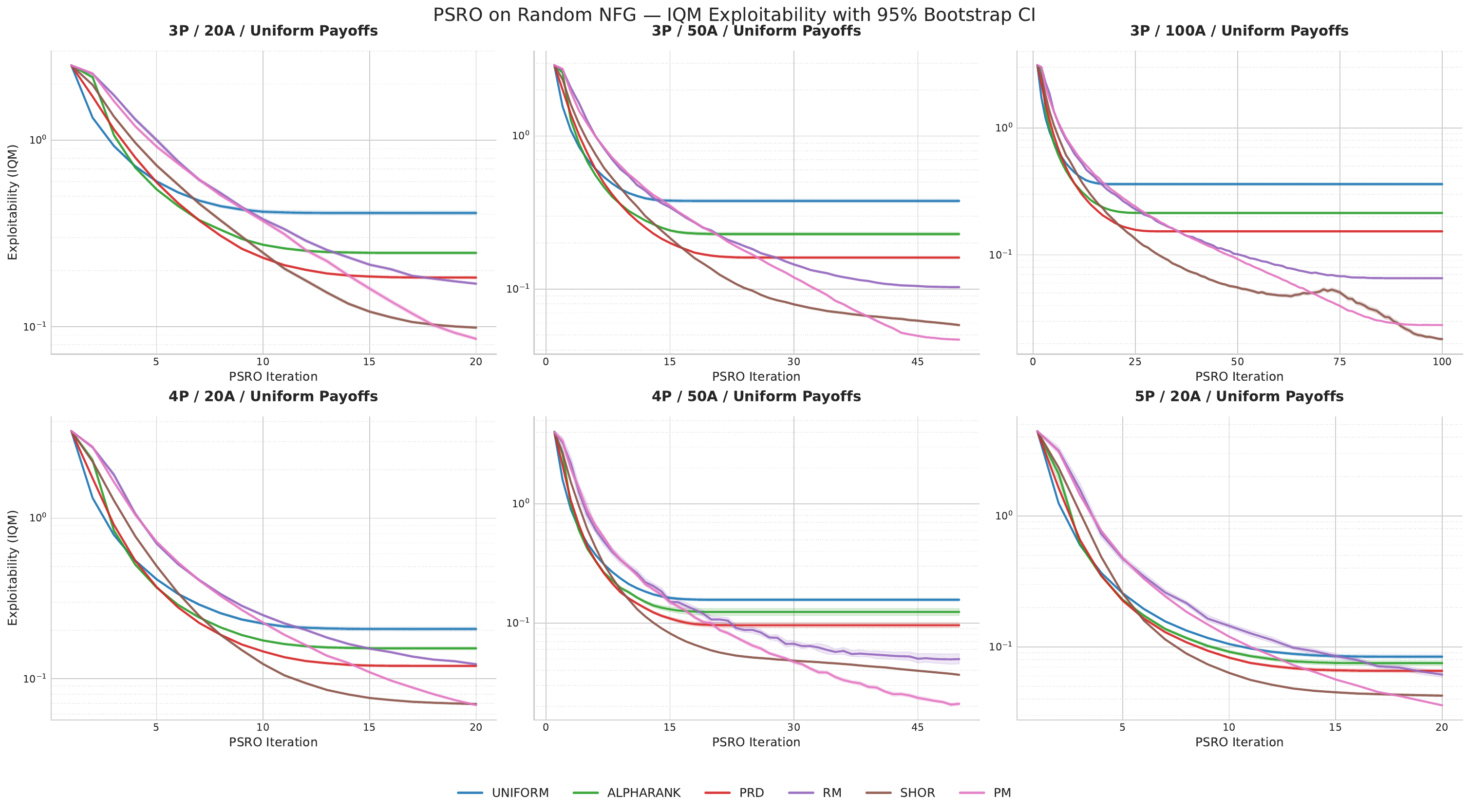}
    \caption{\textbf{PSRO variants performances on Uniform payoff games.}}
    \label{fig:uniform_nfg}
\end{figure*}

\paragraph{Results.} Figures~\ref{fig:gaussian_nfg} and \ref{fig:uniform_nfg} show per-iteration IQM exploitability under Gaussian and Uniform payoffs respectively, across six $(N, A)$ configurations spanning 3-, 4-, and 5-player games with $A$ ranging from 20 to 100 actions. The qualitative pattern is identical across all configurations and both distributions: SHOR-PSRO and PM-PSRO consistently reach lower exploitability than every established baseline, with the gap widening as iteration count grows. Established baselines (Uniform, AlphaRank, PRD, and RM) plateau early — typically before iteration $A/2$ — while SHOR-PSRO and PM-PSRO continue to descend. PM-PSRO begins more slowly during the first $\sim 5$ iterations, since its tangent-projected utility signal requires several iterations to accumulate, but matches or surpasses SHOR-PSRO by mid-run in the larger configurations (3P/50A, 3P/100A, 4P/50A). The 95\% bootstrap CIs are tighter than line width across most of each curve, indicating that the observed gaps are statistically significant rather than artifacts of game-level variance. Insensitivity to the payoff distribution (Gaussian vs. Uniform results are qualitatively identical) suggests that the discovered solvers' advantage is not specific to any particular generative model.

\subsection{Prompts}\label{sec:prompts}

\begin{lstlisting}[caption={Prompt for Evolving CFR}, label=lst:cfr_prompt, language=]
Act as an expert in game theory, multiagent learning, online learning and optimization. Your task is to iteratively improve a new variant of counterfactual regret minimization. The primary goal is to speed up convergence to low-exploitable strategies.

Always adhere to best practices in Python coding.

A key data structure that is used is infostate_node:

```python
@attr.s
class _InfoStateNode(object):
  """An object wrapping values associated to an information state."""
  # The list of the legal actions.
  legal_actions = attr.ib()
  index_in_tabular_policy = attr.ib()
  # Map from information states string representations and actions to the
  # counterfactual regrets, accumulated over the policy iterations
  cumulative_regret = attr.ib(factory=lambda: collections.defaultdict(float))
  # Same as above for the cumulative of the policy probabilities computed
  # during the policy iterations
  cumulative_policy = attr.ib(factory=lambda: collections.defaultdict(float))
```

You are allowed to modify three key components of CFR: (1) how are the regret values accumulated at each infoset (RegretAccumulator) (2) how to obtain a current policy at the current iteration from the current cumulative_regret (PolicyFromRegretAccumulator) and (3) how to accumulate policies across iterations to compute an average policy (PolicyAccumulator).

# Prior programs
Previously we found that the following programs performed well on the task at hand:

{previous_programs}

# Current program
Here is the current program we are trying to improve (you will need to propose a modification to it below):

{code}

# *SEARCH/REPLACE block* Rules:

Every *SEARCH/REPLACE block* must use this format:
1. The opening fence: ```python
2. The start of search block: <<<<<<< SEARCH
3. A contiguous chunk of up to 4 lines to search for in the existing source code
4. The dividing line: =======
5. The lines to replace into the source code
6. The end of the replace block: >>>>>>> REPLACE
7. The closing fence: ```

Every *SEARCH* section must *EXACTLY MATCH* the existing file content, character for character, including all comments, docstrings, etc.

*SEARCH/REPLACE* blocks will replace *all* matching occurrences.
Include enough lines to make the SEARCH blocks uniquely match the lines to change.

Keep *SEARCH/REPLACE* blocks concise.
Break large *SEARCH/REPLACE* blocks into a series of smaller blocks that each change a small portion of the file.
Include just the changing lines, and a few surrounding lines if needed for uniqueness.
Do not include long runs of unchanging lines in *SEARCH/REPLACE* blocks.

To move code within a file, use 2 *SEARCH/REPLACE* blocks: 1 to delete it from its current location, 1 to insert it in the new location.

Make sure that the changes you propose are consistent with each other. For example, if you refer to a new config variable somewhere, you should also propose a change to add that variable.

Example:
```python
<<<<<<< SEARCH
    return total_loss
=======
    # Add sparsity-promoting regularization to the loss.
    total_loss += self.hypers.l1_reg_weight * l1_reg

    return total_loss
{replace}
```
and
```python
<<<<<<< SEARCH
  return hyper.zipit([
=======
  return hyper.zipit([
      hyper.uniform('l1_reg_weight', hyper.interval(0.0, 0.01)),
{replace}
```

{lazy_prompt}
ONLY EVER RETURN CODE IN A *SEARCH/REPLACE BLOCK*!

# Task
{task_instruction} {focus_sentence} {trigger_chain_of_thought}
Describe each change with a *SEARCH/REPLACE block*.
\end{lstlisting}

\begin{lstlisting}[caption={Prompt Config for Evolving CFR}, label=lst:cfr_prompt_config, language=]
{
  "template": {
    "prompt_sampler_type": "TemplatePromptSampler",
    "probability": 1.0,
    "template_path": "meta_cfr_prompt.md",
    "format_args": {
      "lazy_prompt": [
        ["\nYou are diligent and tireless!\nYou NEVER leave comments describing code without implementing it!\nYou always COMPLETELY IMPLEMENT the needed code!", 0.5],
        ["", 0.5]
      ],
      "task_instruction": [
        ["Propose modifications to current program that combine the strengths of all the programs above and achieved high scores on the task.", 0.25],
        ["Suggest an unconventional modification to improve our implementation.", 0.25],
        ["Suggest a new idea to improve the code.", 0.25],
        ["Suggest a new idea to improve the code that is inspired by your expert knowledge of game theory, multiagent reinforcement learning, economics, stochastic processes, machine learning, and optimization.", 0.25]
      ],
      "focus_sentence": [
        ["Focus on simplifying the code, instead of adding new functionality.", 0.2],
        ["", 0.8]
      ],
      "trigger_chain_of_thought": [
        ["\n\nStart with providing a comprehensive explanation for the proposed changes including\n* The specific issue or limitation it addresses.\n* The underlying rationale and expected impact.\n\nYou need to specify this *before providing code*.\n\n", 0.5],
        ["\n\n", 0.5]
      ]
    }
  }
}
\end{lstlisting}

\begin{lstlisting}[caption={Prompt for Evolving PSRO}, label=lst:psro_prompt, language=]
Act as an expert in game theory, multiagent learning, online learning and optimization. Your task is to iteratively improve a variant of Policy-Space Response Oracles (PSRO). The primary goal is to speed up convergence to low-exploitable strategies.

# PSRO Overview

PSRO iteratively builds a population of policies for each player and computes meta-strategies (distributions over policies) to guide training and evaluation.

**Each iteration:**
1. **Empirical game**: Simulate payoffs between all policy combinations to form a game tensor.
2. **Train-time meta-strategy**: Compute a distribution over current policies for each player. This determines what opponents the best-response oracle trains against.
3. **Best response**: Add a new policy for each player that best responds to opponents' train-time meta-strategies.
4. **Eval-time meta-strategy**: Compute a (possibly different) distribution for evaluation, e.g., to measure exploitability.

**Your task**: Improve both the **train-time** and **eval-time** meta-strategy solvers. These serve different purposes: train-time guides population growth, eval-time assesses solution quality.

# Available Utilities

**Best Response**
- `BestResponsePolicy(game, player_id, policy)`: Returns a best-response policy for `player_id` against `policy`.
- Use `.value(game.new_initial_state())` to get the BR value.

**Policy Aggregation**
- `PolicyAggregator(game).aggregate(pids, policy_sets, weights)`: Creates a mixed policy from weighted pure policies.
  - `pids`: `list(range(game.num_players()))`
  - `weights`: `[weights_p0, weights_p1, ...]`, each matching `len(policy_sets[p])`

**Policy Evaluation**
- `expected_game_score.policy_value(state, joint_policy)`: Returns expected payoffs (list, one per player).
  - `joint_policy`: `[policy_p0, policy_p1, ...]`

Always adhere to best practices in Python coding.

# Prior programs
Previously we found that the following programs performed well on the task at hand:

{previous_programs}

# Current program
Here is the current program we are trying to improve (you will need to propose a modification to it below):

{code}

# *SEARCH/REPLACE block* Rules:

Every *SEARCH/REPLACE block* must use this format:
1. The opening fence: ```python
2. The start of search block: <<<<<<< SEARCH
3. A contiguous chunk of up to 4 lines to search for in the existing source code
4. The dividing line: =======
5. The lines to replace into the source code
6. The end of the replace block: >>>>>>> REPLACE
7. The closing fence: ```

Every *SEARCH* section must *EXACTLY MATCH* the existing file content, character for character, including all comments, docstrings, etc.

*SEARCH/REPLACE* blocks will replace *all* matching occurrences.
Include enough lines to make the SEARCH blocks uniquely match the lines to change.

Keep *SEARCH/REPLACE* blocks concise.
Break large *SEARCH/REPLACE* blocks into a series of smaller blocks that each change a small portion of the file.
Include just the changing lines, and a few surrounding lines if needed for uniqueness.
Do not include long runs of unchanging lines in *SEARCH/REPLACE* blocks.

To move code within a file, use 2 *SEARCH/REPLACE* blocks: 1 to delete it from its current location, 1 to insert it in the new location.

Make sure that the changes you propose are consistent with each other. For example, if you refer to a new config variable somewhere, you should also propose a change to add that variable.

Example:
```python
<<<<<<< SEARCH
    return total_loss
=======
    # Add sparsity-promoting regularization to the loss.
    total_loss += self.hypers.l1_reg_weight * l1_reg

    return total_loss
{replace}
```
and
```python
<<<<<<< SEARCH
  return hyper.zipit([
=======
  return hyper.zipit([
      hyper.uniform('l1_reg_weight', hyper.interval(0.0, 0.01)),
{replace}
```

{lazy_prompt}
ONLY EVER RETURN CODE IN A *SEARCH/REPLACE BLOCK*!

# Task
{task_instruction} {focus_sentence} {difficulty_hint} {exploration_nudge} {trigger_chain_of_thought}
Describe each change with a *SEARCH/REPLACE block*.

\end{lstlisting}

\begin{lstlisting}[caption={Prompt Config for Evolving PSRO}, label=lst:psro_prompt_config, language=]
{
  "template": {
    "prompt_sampler_type": "TemplatePromptSampler",
    "probability": 1.0,
    "template_path": "meta_psro_prompt.md",
    "format_args": {
      "lazy_prompt": [
        ["\nYou are diligent and tireless!\nYou NEVER leave comments describing code without implementing it!\nYou always COMPLETELY IMPLEMENT the needed code!", 0.5],
        ["", 0.5]
      ],
      "task_instruction": [
        ["Propose modifications to current program that combine the strengths of all the programs above and achieved high scores on the task.", 0.25],
        ["Suggest an unconventional modification to improve our implementation.", 0.25],
        ["Suggest a new idea to improve the code.", 0.25],
        ["Suggest a new idea to improve the code that is inspired by your expert knowledge of game theory, multiagent reinforcement learning, economics, stochastic processes, machine learning, and optimization.", 0.25]
      ],
      "focus_sentence": [
        ["Focus on simplifying the code, instead of adding new functionality.", 0.2],
        ["", 0.8]
      ],
      "trigger_chain_of_thought": [
        ["\n\nStart with providing a comprehensive explanation for the proposed changes including\n* The specific issue or limitation it addresses.\n* The underlying rationale and expected impact.\n\nYou need to specify this *before providing code*.\n\n", 0.5],
        ["\n\n", 0.5]
      ]
    }
  }
}

\end{lstlisting}

%% file: references.bib
@book{ShohamLeytonBrown2008,
  author    = {Shoham, Yoav and Leyton-Brown, Kevin},
  title     = {Multiagent Systems: Algorithmic, Game-Theoretic, and Logical Foundations},
  publisher = {Cambridge University Press},
  year      = {2008},
  address   = {New York, NY, USA}
}

@inproceedings{Lanctot2017PSRO,
  author    = {Lanctot, Marc and Zambaldi, Vinicius and Gruslys, Audrunas and Lazaridou, Angeliki and Tuyls, Karl and P{\'e}r{\'e}at, Julien and Silver, David and Graepel, Thore},
  title     = {A Unified Game-Theoretic Approach to Multiagent Reinforcement Learning},
  booktitle = {Thirty-First International Conference on Neural Information Processing Systems},
  year      = {2017},
  pages     = {4190--4203},
}

@article{Brown2019Pluribus,
  author    = {Brown, Noam and Sandholm, Tuomas},
  title     = {Superhuman AI for multiplayer poker},
  journal   = {Science},
  volume    = {365},
  number    = {6456},
  pages     = {885--890},
  year      = {2019},
  doi       = {10.1126/science.aay2400},
  publisher = {American Association for the Advancement of Science}
}

@inproceedings{brown2019solving,
  title={Solving imperfect-information games via discounted regret minimization},
  author={Brown, Noam and Sandholm, Tuomas},
  booktitle={Thirty-Third AAAI Conference on Artificial Intelligence},
  volume={33},
  pages={1829--1836},
  year={2019}
}

@inproceedings{mcmahan2003planning,
  title={Planning in the presence of cost functions controlled by an adversary},
  author={McMahan, H Brendan and Gordon, Geoffrey J and Blum, Avrim},
  booktitle={Twentieth International Conference on Machine Learning},
  pages={536--543},
  year={2003}
}

@article{Vinyals2019AlphaStar,
  author    = {Vinyals, Oriol and Babuschkin, Igor and Czarnecki, Wojciech M. and Mathieu, Michaël and Dudzik, Andrew and others},
  title     = {Grandmaster level in StarCraft II using multi-agent reinforcement learning},
  journal   = {Nature},
  volume    = {575},
  number    = {7782},
  pages     = {350--354},
  year      = {2019},
  publisher = {Nature Publishing Group},
}

@inproceedings{Zinkevich2007CFR,
  author    = {Zinkevich, Martin and Johanson, Michael and Bowling, Michael and Piccione, Carmelo},
  title     = {Regret Minimization in Games with Incomplete Information},
  booktitle = {Twenty-First International Conference on Neural Information Processing Systems},
  year      = {2007},
  pages     = {1729--1736},
}

@inproceedings{agarwal2021deep,
  title={Deep reinforcement learning at the edge of the statistical precipice},
  author={Agarwal, Rishabh and Schwarzer, Max and Castro, Pablo Samuel and Courville, Aaron C and Bellemare, Marc},
  booktitle={Thirty-Fifth Conference on Neural Information Processing Systems},
  volume={34},
  pages={29304--29320},
  year={2021}
}

@inproceedings{lu2024discovering,
  title={Discovering preference optimization algorithms with and for large language models},
  author={Lu, Chris and Holt, Samuel and Fanconi, Claudio and Chan, Alex J and Foerster, Jakob and van der Schaar, Mihaela and Lange, Robert T},
  booktitle={Thirty-Sixth Conference on Neural Information Processing Systems},
  volume={37},
  pages={86528--86573},
  year={2024}
}

@article{sandholm2008projection,
  title={The projection dynamic and the replicator dynamic},
  author={Sandholm, William H and Dokumac{\i}, Emin and Lahkar, Ratul},
  journal={Games and Economic Behavior},
  volume={64},
  number={2},
  pages={666--683},
  year={2008},
  publisher={Elsevier}
}

@article{tammelin2014solving,
  title={Solving large imperfect information games using CFR+},
  author={Tammelin, Oskari},
  journal={arXiv preprint arXiv:1407.5042},
  year={2014}
}

@article{novikov2025alphaevolve,
  title={AlphaEvolve: A coding agent for scientific and algorithmic discovery},
  author={Novikov, Alexander and V{\~u}, Ng{\^a}n and Eisenberger, Marvin and Dupont, Emilien and Huang, Po-Sen and Wagner, Adam Zsolt and Shirobokov, Sergey and Kozlovskii, Borislav and Ruiz, Francisco JR and Mehrabian, Abbas and others},
  journal={arXiv preprint arXiv:2506.13131},
  year={2025}
}

@article{lanctot2019openspiel,
  title={OpenSpiel: A framework for reinforcement learning in games},
  author={Lanctot, Marc and Lockhart, Edward and Lespiau, Jean-Baptiste and Zambaldi, Vinicius and Upadhyay, Satyaki and P{\'e}rolat, Julien and Srinivasan, Sriram and Timbers, Finbarr and Tuyls, Karl and Omidshafiei, Shayegan and others},
  journal={arXiv preprint arXiv:1908.09453},
  year={2019}
}

@inproceedings{bighashdel2024policy,
  title={Policy Space Response Oracles: A Survey},
  author={Bighashdel, Ariyan and Wang, Yongzhao and McAleer, Stephen and Savani, Rahul and Oliehoek, Frans A},
  booktitle={Thirty-Third International Joint Conference on Artificial Intelligence (IJCAI-24) Survey Track},
  year={2024}
}

@inproceedings{farina2021faster,
  title={Faster game solving via predictive blackwell approachability: Connecting regret matching and mirror descent},
  author={Farina, Gabriele and Kroer, Christian and Sandholm, Tuomas},
  booktitle={Thirty-Fifth AAAI Conference on Artificial Intelligence},
  volume={35},
  pages={5363--5371},
  year={2021}
}

@inproceedings{xu2024minimizing,
  title={Minimizing weighted counterfactual regret with optimistic online mirror descent},
  author={Xu, Hang and Li, Kai and Liu, Bingyun and Fu, Haobo and Fu, Qiang and Xing, Junliang and Cheng, Jian},
  booktitle={Thirty-Third International Joint Conference on Artificial Intelligence},
  pages={5272--5280},
  year={2024}
}

@inproceedings{brown2016strategy,
  title={Strategy-based warm starting for regret minimization in games},
  author={Brown, Noam and Sandholm, Tuomas},
  booktitle={Thirtieth AAAI Conference on Artificial Intelligence},
  volume={30},
  year={2016}
}

@inproceedings{mullergeneralized,
  title={A Generalized Training Approach for Multiagent Learning},
  author={Muller, Paul and Omidshafiei, Shayegan and Rowland, Mark and Tuyls, Karl and Perolat, Julien and Liu, Siqi and Hennes, Daniel and Marris, Luke and Lanctot, Marc and Hughes, Edward and others},
  booktitle={Ninth International Conference on Learning Representations},
  year = {2020},
}

@inproceedings{zhang2024faster,
  title={Faster game solving via hyperparameter schedules},
  author={Zhang, Naifeng and McAleer, Stephen and Sandholm, Tuomas},
  booktitle={Fortieth AAAI Conference on Artificial Intelligence},
  year={2026}
}

@article{wellman2025empirical,
  title={Empirical game theoretic analysis: A survey},
  author={Wellman, Michael P and Tuyls, Karl and Greenwald, Amy},
  journal={Journal of Artificial Intelligence Research},
  volume={82},
  pages={1017--1076},
  year={2025}
}

@inproceedings{zhang2024exponential,
  title={Exponential lower bounds on the double oracle algorithm in zero-sum games},
  author={Zhang, Brian Hu and Sandholm, Tuomas},
  booktitle={Thirty-Third International Joint Conference on Artificial Intelligence},
  pages={3032--3039},
  year={2024}
}

@article{comanici2025gemini,
  title={Gemini 2.5: Pushing the frontier with advanced reasoning, multimodality, long context, and next generation agentic capabilities},
  author={Comanici, Gheorghe and Bieber, Eric and Schaekermann, Mike and Pasupat, Ice and Sachdeva, Noveen and Dhillon, Inderjit and Blistein, Marcel and Ram, Ori and Zhang, Dan and Rosen, Evan and others},
  journal={arXiv preprint arXiv:2507.06261},
  year={2025}
}

@inproceedings{xu2022autocfr,
  title={Autocfr: Learning to design counterfactual regret minimization algorithms},
  author={Xu, Hang and Li, Kai and Fu, Haobo and Fu, Qiang and Xing, Junliang},
  booktitle={Thirty-Sixth AAAI Conference on Artificial Intelligence},
  pages={5244--5251},
  year={2022}
}

@inproceedings{xu2024dynamic,
  title={Dynamic discounted counterfactual regret minimization},
  author={Xu, Hang and Li, Kai and Fu, Haobo and Fu, Qiang and Xing, Junliang and Cheng, Jian},
  booktitle={Twelfth International Conference on Learning Representations},
  year={2024}
}

@inproceedings{fengneural,
  title={Neural Auto-Curricula in Two-Player Zero-Sum Games},
  author={Feng, Xidong and Slumbers, Oliver and Wan, Ziyu and Liu, Bo and McAleer, Stephen Marcus and Wen, Ying and Wang, Jun and Yang, Yaodong},
  booktitle={Thirty-Fifth International Conference on Neural Information Processing Systems},
  year={2021},
}

@inproceedings{real2020automl,
  title={AutoML-Zero: Evolving Machine Learning Algorithms From Scratch},
  author={Real, Esteban and Liang, Chen and So, David and Le, Quoc V},
  booktitle={Thirty-Seventh International Conference on Machine Learning},
  pages={8007--8019},
  year={2020},
}

@inproceedings{xu2018meta,
  title={Meta-Gradient Reinforcement Learning},
  author={Xu, Zhongwen and van Hasselt, Hado and Silver, David},
  booktitle={Thirty-Second International Conference on Neural Information Processing Systems},
  volume={31},
  year={2018}
}

@inproceedings{metz2019understanding,
  title={Understanding and Correcting Pathologies in the Training of Learned Optimizers},
  author={Metz, Luke and Maheswaranathan, Niru and Nixon, Jeremy and Freeman, C Daniel and Sohl-Dickstein, Jascha},
  booktitle={Thirty-Sixth International Conference on Machine Learning},
  pages={4556--4565},
  year={2019},
}

@article{georgiev2025mathematical,
  title={Mathematical exploration and discovery at scale},
  author={Georgiev, Bogdan and G{\'o}mez-Serrano, Javier and Tao, Terence and Wagner, Adam Zsolt},
  journal={arXiv preprint arXiv:2511.02864},
  year={2025}
}

@article{oh2025discovering,
  title={Discovering state-of-the-art reinforcement learning algorithms},
  author={Oh, Junhyuk and Farquhar, Greg and Kemaev, Iurii and Calian, Dan A and Hessel, Matteo and Zintgraf, Luisa and Singh, Satinder and Van Hasselt, Hado and Silver, David},
  journal={Nature},
  pages={1--2},
  year={2025},
  publisher={Nature Publishing Group UK London}
}

@article{nagda2025reinforced,
  title={Reinforced Generation of Combinatorial Structures: Hardness of Approximation},
  author={Nagda, Ansh and Raghavan, Prabhakar and Thakurta, Abhradeep},
  journal={arXiv preprint arXiv:2509.18057},
  year={2025}
}

@inproceedings{chen2023symbolic,
  title={Symbolic discovery of optimization algorithms},
  author={Chen, Xiangning and Liang, Chen and Huang, Da and Real, Esteban and Wang, Kaiyuan and Pham, Hieu and Dong, Xuanyi and Luong, Thang and Hsieh, Cho-Jui and Lu, Yifeng and others},
  booktitle={Thirty-Seventh International Conference on  neural information processing systems},
  pages={49205--49233},
  year={2023}
}

@inproceedings{coevolving,
  title={Evolving Reinforcement Learning Algorithms},
  author={Co-Reyes, John D and Miao, Yingjie and Peng, Daiyi and Real, Esteban and Le, Quoc V and Levine, Sergey and Lee, Honglak and Faust, Aleksandra},
  booktitle={Ninth International Conference on Learning Representations},
  year={2021},
}

@inproceedings{Sychrovsky24,
  title={Learning to Not Regret},
  author={David Sychrovsk'{y} and Michal \v{S}ustr and Elnaz Davoodi
and Michael Bowling and Marc Lanctot and Martin Schmid},
  booktitle = {Thirty-Eighth AAAI Conference on Artificial Intelligence},
  year = {2024},
}

@article{kuhn1950simplified,
  title={Simplified two-person poker},
  author={Kuhn, Harold W.},
  journal={Contributions to the Theory of Games},
  pages={97--103},
  year={1950},
  publisher={Princeton University Press}
}

@inproceedings{southey2005bayes,
  title={Bayes' bluff: opponent modelling in poker},
  author={Southey, Finnegan and Bowling, Michael and Larson, Bryce and Piccione, Carmelo and Burch, Neil and Billings, Darse and Rayner, Chris},
  booktitle={Twenty-First Conference on Uncertainty in Artificial Intelligence},
  pages={550--558},
  year={2005}
}

@article{ross1971goofspiel,
  title={Goofspiel—the game of pure strategy},
  author={Ross, Sheldon M},
  journal={Journal of Applied Probability},
  volume={8},
  number={3},
  pages={621--625},
  year={1971},
  publisher={Cambridge University Press}
}

@inproceedings{lisy2015online,
  title={Online Monte-Carlo counterfactual regret minimization for search in imperfect information games},
  author={Lis{\`y}, Viliam and Lanctot, Marc and Bowling, Michael H},
  booktitle={Fourteenth International Conference on Autonomous Agents and Multiagent Systems},
  pages={27--36},
  year={2015}
}

@article{borel1953theory,
  title={The theory of play and integral equations with skew symmetric kernels},
  author={Borel, Emile},
  journal={Econometrica: journal of the Econometric Society},
  pages={97--100},
  year={1953},
  publisher={JSTOR}
}

@inproceedings{farina2019correlation,
  title={Correlation in extensive-form games: Saddle-point formulation and benchmarks},
  author={Farina, Gabriele and Ling, Chun Kai and Fang, Fei and Sandholm, Tuomas},
  booktitle={Thirty-third International Conference on Neural Information Processing Systems},
  volume={32},
  year={2019}
}

@article{lahkar2008projection,
  title={The projection dynamic and the geometry of population games},
  author={Lahkar, Ratul and Sandholm, William H},
  journal={Games and Economic Behavior},
  volume={64},
  number={2},
  pages={565--590},
  year={2008},
  publisher={Elsevier}
}

@inproceedings{brown2019deep,
  title={Deep counterfactual regret minimization},
  author={Brown, Noam and Lerer, Adam and Gross, Sam and Sandholm, Tuomas},
  booktitle={Thirty-Sixth International Conference on Machine Learning},
  pages={793--802},
  year={2019}
}
